\documentclass{article}

\usepackage{arxiv}

\usepackage{authblk}
\usepackage{graphicx} 
\usepackage{booktabs}
\usepackage{multirow}
\usepackage{amsmath}
\usepackage{caption}
\usepackage{subcaption}
\usepackage{float}
\usepackage{placeins}
\usepackage[printonlyused, nohyperlinks]{acronym}
\usepackage{tcolorbox}
\tcbuselibrary{breakable}

\usepackage{listings}
\usepackage{xcolor}

\definecolor{codegreen}{rgb}{0,0.6,0}
\definecolor{codered}{rgb}{1,0,0}
\definecolor{codegray}{rgb}{0.5,0.5,0.5}
\definecolor{codepurple}{rgb}{0.58,0,0.82}
\definecolor{backcolour}{rgb}{0.95,0.95,0.92}

\lstdefinestyle{mystyle}{
  backgroundcolor=\color{backcolour},
  commentstyle=\color{codegreen},
  keywordstyle=\color{magenta},
  numberstyle=\tiny\color{codegray},
  stringstyle=\color{codepurple},
  basicstyle=\ttfamily\footnotesize,
  breakatwhitespace=false,         
  breaklines=true,                 
  captionpos=b,                    
  keepspaces=true,                 
  numbers=left,                    
  numbersep=5pt,                  
  showspaces=false,                
  showstringspaces=false,
  showtabs=false,                  
  tabsize=2,
  escapeinside={(*@}{@*)}
}

\usepackage{tikz}
\def\checkmark{\tikz\fill[scale=0.4](0,.35) -- (.25,0) -- (1,.7) -- (.25,.15) -- cycle;}

\usepackage{biblatex}
\usepackage{array}
\newcolumntype{P}[1]{>{\centering\arraybackslash}p{#1}}

\title{PRISM: A Multi-Modal Generative Foundation Model for Slide-Level Histopathology}

\author[1]{George Shaikovski\textsuperscript{\textdagger}}
\author[1]{Adam Casson\textsuperscript{\textdagger}}
\author[2]{Kristen Severson}
\author[2]{Eric Zimmermann}
\author[1]{Yi Kan Wang}
\author[1]{Jeremy D. Kunz}
\author[1]{Juan A. Retamero}
\author[1]{Gerard Oakley}
\author[1]{David Klimstra}
\author[1,3]{Christopher Kanan}
\author[1,4]{Matthew Hanna}
\author[1]{Michal Zelechowski}
\author[1]{Julian Viret}
\author[2]{Neil Tenenholtz}
\author[2]{James Hall}
\author[2]{Nicol\`o Fusi}
\author[1]{Razik Yousfi}
\author[1]{Peter Hamilton}
\author[1]{William A. Moye}
\author[1]{Eugene Vorontsov}
\author[1]{Siqi Liu\textsuperscript{\textdaggerdbl}}
\author[1]{Thomas J. Fuchs}

\affil[1]{Paige, NYC, NY United States}
\affil[2]{Microsoft Research, Cambridge, MA United States}
\affil[3]{University of Rochester, Rochester, NY United States}
\affil[4]{Memorial Sloan Kettering Cancer Center, NYC, NY United States}

\begin{document}

\date{}
\maketitle

\def\thefootnote{\textdaggerdbl}\footnotetext{ Corresponding author. siqi.liu AT paige DOT ai\\}\def\thefootnote{\arabic{footnote}}

\def\thefootnote{\textdagger}\footnotetext{These authors contributed equally to this work.\\}\def\thefootnote{\arabic{footnote}}

\begin{abstract}

Foundation models in computational pathology promise to unlock the development of new clinical decision support systems and models for precision medicine. However, there is a mismatch between most clinical analysis, which is defined at the level of one or more whole slide images, and foundation models to date, which process the thousands of image tiles contained in a whole slide image separately. The requirement to train a network to aggregate information across a large number of tiles in multiple whole slide images limits these models' impact. In this work, we present a slide-level foundation model for H\&E-stained histopathology, \textit{PRISM}, that builds on Virchow tile embeddings and leverages clinical report text for pre-training. Using the tile embeddings, PRISM produces slide-level embeddings with the ability to generate clinical reports, resulting in several modes of use. Using text prompts, PRISM achieves zero-shot cancer detection and sub-typing performance approaching and surpassing that of a supervised aggregator model. Using the slide embeddings with linear classifiers, PRISM surpasses supervised aggregator models. Furthermore, we demonstrate that fine-tuning of the PRISM slide encoder yields label-efficient training for biomarker prediction, a task that typically suffers from low availability of training data; an aggregator initialized with PRISM and trained on as little as 10\% of the training data can outperform a supervised baseline that uses all of the data.

\end{abstract}

\section{Introduction}

Recent progress in computational pathology, fueled by deep learning and extensive pathology image datasets, has facilitated the creation of clinical decision support systems with the primary goal of detecting cancer in \acp{WSI}~\cite{campanella2019clinical, raciti2020novel, da2021independent,perincheri2021independent}. Already, computational pathology tools are available for use in clinical practice~\cite{raciti2023clinical}. Motivated by this success, the field has grown to encompass a broad range of clinically-relevant tasks including cancer sub-typing, microenvironment characterisation, biomarker detection~\cite{coudray2018classification, wagner2023fully}, treatment response, and overall survival prediction. This research has the potential to revolutionize the field of pathology and, ultimately, clinical oncology practices.

\begin{figure*}[!t]
    \centering
    \includegraphics[width=\linewidth]{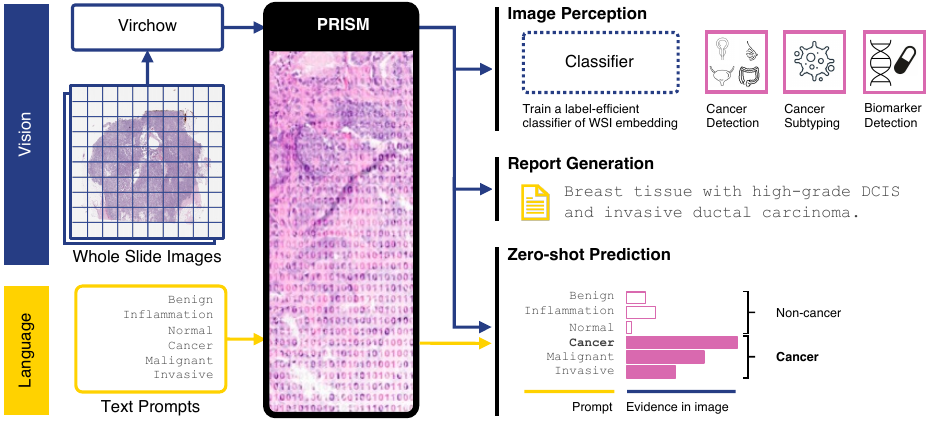}
    \caption{An overview of the capabilities enabled by the slide-level foundation model (PRISM), built on Virchow~\cite{vorontsov2024virchow} tile embeddings. Whereas Virchow produces an embedding for each foreground tile of a set of whole slide images, PRISM aggregates these embeddings into a single slide embedding that can be used for image perception by training a linear classifier for downstream tasks including cancer detection, cancer sub-typing, and biomarker detection. Optionally, the model can be fine-tuned for the classification task. Language-enabled capabilities of PRISM include the training-free ``zero-shot'' prediction via text prompting, and generation of interpretable free text clinical reports.}
    \label{fig:teaser}
\end{figure*}

To build on this progress, recent efforts have focused on creating foundation models capable of operating on tile images (small sub-regions of the \ac{WSI}), trained with self-supervised learning on datasets ranging from tens of thousands to millions of \acp{WSI}~\cite{wang2022transformer,filiot2023scaling,chen_general-purpose_2023,vorontsov2024virchow}.
These models are constrained to tile images because \acp{WSI} are gigapixels in size, or approximately five orders of magnitude larger than a typical natural image.
At such a scale, hardware constraints limit the applicability of many deep learning architectures. 
While foundation models can act as building blocks for a wide range of downstream tasks, most ground truth labels obtainable from clinical databases without manual annotations---such as diagnosis, biomarker status, treatment response, or survival---are linked to a whole slide image or a collection of \acp{WSI} created from the same tissue sample.
Therefore, these foundation models must be paired with \ac{MIL} aggregator models ~\cite{campanella2019clinical,ilse2018attention,li2021dual,lu2021data,shao2021transmil} to map the tile-level embeddings to slide or specimen-level ground truth labels.
Although tile-level foundation models have been pre-trained, aggregator models are typically trained from scratch on each slide-level task. Given a large number of high-dimensional embeddings per \ac{WSI}, these aggregator models are prone to overfitting, particularly when trained on small datasets.
We believe that large-scale pre-training on whole slide images with natural language supervision can tackle the challenge of high dimensionality and limited label availability, and create a generally-applicable slide-level foundation model for pathology.

In this work, we present a multi-modal slide-level foundation model named \textit{PRISM}, for Pathology Report and Image Summarization Model.
To pre-train the model, we use clinical report data as \ac{WSI}-level supervisory signal. 
We show that the proposed pre-training method improves performance on downstream tasks, compared to fully-supervised models.
Importantly, we demonstrate that pre-training can also benefit specific downstream tasks not covered by the report text.
As clinical reports can be obtained without further expensive manual annotation, the proposed pre-training approach can be easily scaled to millions of \acp{WSI}.
PRISM is pre-trained using 587 thousand \acp{WSI} and 195 thousand associated clinical text reports.
The model is capable of generating text-based diagnosis reports for \acp{WSI}. 
While the primary objective of this work is not to produce clinically usable reports, we demonstrate that this pre-training method produces the model that can accurately identify various \ac{WSI} features and match the slides to a correct prompt (zero-shot prediction) or generate a text caption describing the features (report generation), without further training. These include tissue types, the presence and sub-types of cancer, non-neoplastic patterns such as Crohn's disease, inflammation, polyps, and other diagnostic information.
Besides report generation and zero-shot prediction, a \textit{slide embedding} can be obtained from the slide encoder of PRISM for \ac{WSI}-level linear classification with specific ground truth labels; alternatively, the model weights can also be further fine-tuned on smaller datasets (image perception).
See Fig~\ref{fig:teaser} for the overview of PRISM's capabilities.

We evaluate PRISM on cancer detection, tissue sub-typing, and biomarker prediction tasks using zero-shot classification, linear probing, and fine-tuning. Zero-shot cancer classification and sub-typing performance approaches or surpasses that of the supervised cancer detection baselines \cite{vorontsov2024virchow}. Fine-tuned on various biomarkers, PRISM shows superior performance in the low data regime compared to the supervised baselines, and with limited training data, often outperforms supervised training on the full dataset. We also perform a qualitative evaluation of randomly selected generated reports to contextualize PRISM performance. Our contributions can be summarised as follows:
\begin{enumerate}
\item A generative slide-level pathology foundation model pre-trained with clinical report supervision.
\item A memory-efficient training methodology to enable \ac{WSI}-level pre-training.
\item Evaluation of the model's diagnostic capabilities and label efficiency on clinically relevant tasks.
\end{enumerate}

\section{Related Work}

\textbf{Tile-level foundation models.} Vision-based foundation models have recently emerged in computational pathology for \ac{WSI} tiles. These self-supervised models are trained with contrastive learning methods. Initial works employed public datasets, including \ac{TCGA}, with tiles extracted from on the order of 10,000 \acp{WSI} to train models with up to 307 million parameters, based on both convolutional and transformer architectures~\cite{wang2022transformer,ciga2022self,azizi2023robust,filiot2023scaling,kang2023benchmarking}. More recent works trained a \ac{ViT}~\cite{dosovitskiy2020image} (22 million to 632 million parameters) on larger proprietary datasets, including 100,000 to 1.5 million \acp{WSI}~\cite{campanella2023computational,dippel2024rudolfv,chen_general-purpose_2023,vorontsov2024virchow}. The Uni work~\cite{chen_general-purpose_2023} demonstrates that model performance scales with dataset size. Furthermore, Virchow~\cite{vorontsov2024virchow} demonstrates that performance can be further improved by scaling both the model and dataset sizes, approaching clinically relevant cancer prediction performance for specimen-level aggregators trained on Virchow tile embeddings. Nevertheless, the need to train an aggregator network makes it difficult to substantially improve performance on low data tasks like biomarker prediction. This motivates pre-training of the aggregator network in a more complete foundation model.

Some recent works incorporate language for pre-training tile-level models~\cite{gamper2021multiple, huang2023visual, lu2023towards, lu2023visual}. Gamper et al.~\cite{gamper2021multiple} encode all tissue tiles with a \ac{CNN} and pass to a transformer for autoregressive caption decoding. Later methods build on the CLIP~\cite{radford2021learning} approach which introduces a contrastive objective between the representations of a tile and its corresponding description. PLIP~\cite{huang2023visual} extends this to pathology images collected from educational materials. CONCH~\cite{lu2023towards} adopts the CoCa framework~\cite{yu2022coca} that adds a generative objective to CLIP, allowing the model to learn autoregressive caption generation from tiles. Both PLIP and CONCH can be used at the slide-level by simple top-k max-pooling (or more complex regional graph-based pooling) of tile-level predictions. Our work also builds on the CoCa framework but takes it further by training an aggregator at the specimen level (including one or more \acp{WSI}) to predict summarized clinical reports. The Perceiver architecture~\cite{jaegle2021perceiver} is used to make it possible to aggregate over hundreds of thousands of tiles.

\textbf{Slide-level foundation models.} Most work on pre-trained pathology models has focused on tile-level representations. There are two notable exceptions: \ac{HIPT}~\cite{chen2022scaling} and LongViT~\cite{wang2023image}. \ac{HIPT} proposes a hierarchical two-stage model where a \ac{ViT} learns tile representations in the first stage and another \ac{ViT} aggregates the class token representations of regional tiles. Whereas \ac{HIPT} only summarizes local regions, LongViT summarizes a full \ac{WSI} using LongNet~\cite{ding2023longnet} attention to allow for very long context lengths. This relies on dilated attention in each head looking at a unique subset of a sequence of tiles. These works differ from the proposed method in that they rely on self-supervised objectives that lack clinical-report-based supervision. We posit that this supervision may help to disentangle more clinically relevant features. Clinical report generation in the proposed method also enables zero-shot prediction.

\textbf{Pathology report generation.} Early pathology report generation work for \ac{WSI} used \ac{CNN} tile encoding and \ac{RNN} caption decoding~\cite{zhang2020evaluating, tsuneki2022inference, qin2023whole}. Later work improved the spatial context inferred from the image by using \ac{HIPT} instead of a \ac{CNN}. These methods relied on randomly subsampling the tiles in \ac{WSI} and using memory-efficient \acp{RNN} which are slower to train and harder to parallelize at scale than the transformers commonly used in state of the art language models.

\section{Slide-Level Foundation Model}

\begin{figure*}[!t]
    \centering
    \includegraphics[width=1.0\linewidth]{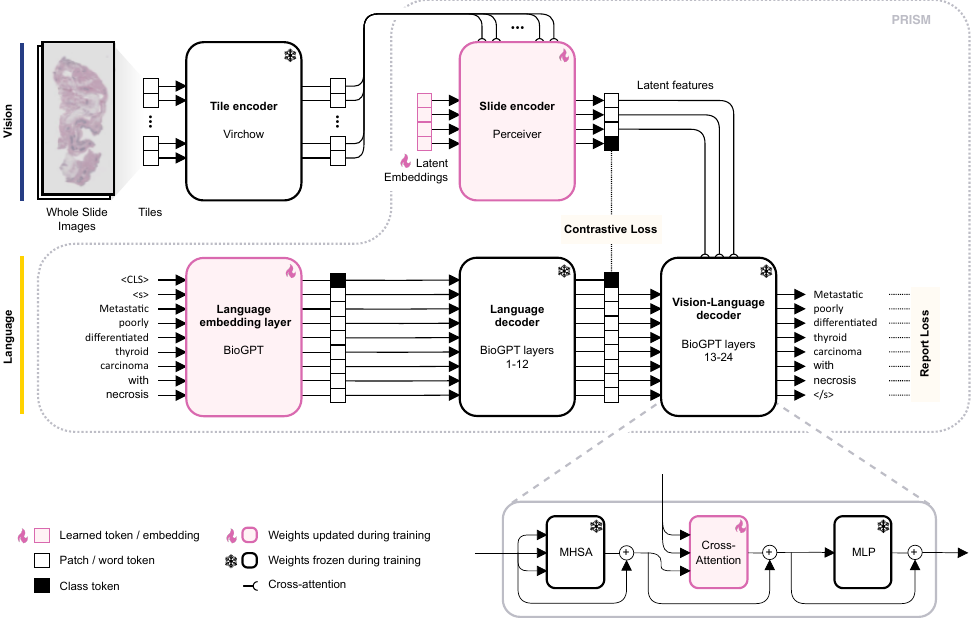}
    \caption{The training methodology for the slide-level foundation model (PRISM). All trained weights are initialized to random values except for the BioGPT word embeddings. Whole slide images and clinical report latent embeddings are aligned with a contrastive loss. Report generation is trained with a generative loss using teacher forcing. Layers 13-24 of the BioGPT decoder are modified to cross-attend to vision embeddings.}
    \label{fig:arch}
\end{figure*}

This work combines natural-language supervision and cross-attention-based resampling to overcome the computational challenges of training with whole slide images while effectively directing the learning algorithm to find important, generalizable features in the slides. PRISM, the slide-level foundation model, contains two components: a slide encoder which leverages a Perceiver network~\cite{jaegle2021perceiver}, and a language decoder which leverages the BioGPT language model~\cite{luo2022biogpt}. Using paired samples of clinical reports, which are rewritten using GPT-4~\cite{achiam2023gpt}, and \acp{WSI}, which are pre-processed to create a sequence of tile-level embeddings using our Virchow foundation model~\cite{vorontsov2024virchow}, PRISM is trained following the CoCa methodology~\cite{yu2022coca} which uses two objectives: (1) alignment of the the encoded report embedding from the pre-trained BioGPT~\cite{luo2022biogpt} language decoder with the slide latent embedding, and (2) prediction of the generated caption by BioGPT to match the ground-truth rewritten report.
The full pre-training framework is shown in Fig.~\ref{fig:arch} and described in detail below.
After pre-training, the combined Virchow-PRISM model can be used to produce a single embedding of one or more \acp{WSI}, and generate reports to describe histological features in the slides.

\textbf{Pre-processing of clinical reports.}
A training sample for PRISM contains multiple whole slide images grouped into a specimen, and the associated natural-language clinical report.
The reports typically contain a mixture of pre-defined structured fields (synoptic reports) and free-text fields, although the formats can vary significantly, particularly for different sites of tissue samples.
The reports are first processed to extract the clinical diagnosis section using heuristic text matching rules.
If available, \ac{IHC} or molecular test results are appended to the diagnosis section. 

The concatenated reports are rewritten into concise text summaries using GPT-4~\cite{achiam2023gpt} to reduce their length, and increase the density of relevant diagnostic information per report.
The rewritten summaries may include, but are not limited to, sites of tissue sample, presence of cancer, its sub-type and grade if available, non-cancerous tissue, \ac{IHC} status, and potential molecular testing results.
Each specimen report is rewritten five times by GPT-4 (Section~\ref{sec:apx_gpt4_rewrites}) to be sampled randomly during training to prevent the network from memorising specimen-to-report associations.

\textbf{Tile embeddings.} 
During training and inference, each \ac{WSI} is divided into a uniform grid of 224$\times$224 pixel image tiles at a 20$\times$ magnification level, corresponding to a resolution of 0.5 µm/pixel.
Tiles with at least 25\% tissue by area are used during training and the rest are discarded. To estimate the amount of tissue in a tile, each \ac{WSI} is downsampled 16$\times$ with bilinear interpolation and every pixel of the downsampled image is evaluated as to whether its hue, saturation, and value are within [90, 180], [8, 255], and [103, 255], respectively.
These tiles are subsequently processed with Virchow~\cite{vorontsov2024virchow}, a tile-level foundation model, to produce tile embeddings.
The Virchow model is a 632M parameter vision transformer (\acs{ViT}-H/14)~\cite{dosovitskiy2020image} trained using DINOv2~\cite{oquab2023dinov2} on over 2B image tiles extracted from 1.5 million \ac{HE}-stained \acp{WSI}.
This large scale pre-training leads to embeddings which can capture the large diversity of tissue and morphological patterns~\cite{vorontsov2024virchow}.
The resulting embedding is a 2560-dimensional vector constructed by concatenating the output class token (1280 dimensions) with the mean across all output patch tokens (1280 dimensions).

\textbf{\acs{WSI} embeddings.} 
Encoding whole slide images into a single global representation is highly compressive, given that each specimen can consist of hundreds of thousands of tiles.
We use the memory-efficient Perceiver network~\cite{jaegle2021perceiver} to aggregate the long sequence of specimen tile embeddings into a much smaller fixed-length set of latent features using trainable latent embeddings (``latents'') and an asymmetric cross-attention mechanism.

The Perceiver slide aggregator consists of 8 blocks (see Section~\ref{sec:apx_perceiver} for implementation details).
Each block has a cross-attention module followed by a 6-layer latent self-attention transformer~\cite{vaswani2017attention}.
A cross-attention module takes tile embeddings and latents as inputs, and outputs latents.
A latent transformer processes the latents returned by the preceding cross-attention module.
The weights of the cross-attention modules from the 2nd through the 8th Perceiver blocks are shared. Similarly, the weights of the latent transformers from the 1st through the 8th Perceiver blocks are also shared. The context key-value pairs computed from the input embeddings are cached between the weight-sharing cross-attention modules to decrease memory consumption.

In this study, the Perceiver network uses 513 learned latents with 1280 dimensions.
After being processed by the network, 512 output latents are used as latent context features in the language decoder to generate reports.
The remaining output latent is the \textit{slide embedding} used for the contrastive loss described below.
We found that using an extra slide latent, instead of mean-pooling introduced by Jaegle et al.~\cite{jaegle2021perceiver}, produced more accurate reports.
The slide embedding serves as a global slide-level or specimen-level representation in downstream applications.

\textbf{Text embeddings and report generation.}
We use the pre-trained BioGPT~\cite{luo2022biogpt} as the language network for PRISM. BioGPT is a decoder-only language model tailored to biology applications and is based on the 345M parameter variant of the GPT-2 architecture~\cite{radford2019language}. The total BioGPT vocabulary is 42,384 tokens. We add a class token ($<$CLS$>$ in Figure~\ref{fig:arch}) to the vocabulary to produce a global clinical report representation. We also add cross-attention modules for context-aware report generation using latent features from the Perceiver model.
We use the BioGPT tokenizer and the embedding layer to encode the training text prompt as a sequence of word embeddings with positional information.

Following Yu et al.~\cite{yu2022coca}, we split the decoder into uni-modal and multi-modal parts.
The uni-modal part includes the layers 1 through 12.
We append the class ($<$CLS$>$) token to the input sequence.
Self-attention over language tokens uses causal masking to prevent look-ahead when generating the next tokens, while the $<$CLS$>$ token attends to all language tokens and to itself.
After layer 12 we separate the $<$CLS$>$ token embedding from the sequence.
This embedding represents the ground truth clinical report and is used in computing the contrastive loss via alignment with the corresponding slide embedding elaborated below.
Further in the text we refer to it as report embedding or prompt embedding depending on the model application context.
The remaining language tokens proceed to the multi-modal part of the decoder, which are the layers 13 through 24.
Each layer is augmented with a cross-attention module, inserted between the self-attention module and the MLP network, which take 512 latent features from the Perceiver as key-value context.

We initialize the network with pre-trained BioGPT weights and freeze all of them except the initial word embedding layer.
The initial class token embedding and the weights of cross-attention modules in the multi-modal part are initialized to random values from the Normal distribution.

\textbf{Training objective.}
Following CoCa~\cite{yu2022coca}, the model is trained using contrastive and generative objectives.
To compute the contrastive loss, the slide embedding from the Perceiver and the paired report embedding from BioGPT are projected using a linear layer with 5120 output dimensions and normalized with the Euclidean $\ell_2$ norm.
The objective maximizes the cosine similarity between the paired projections, while minimizing the similarity of the unmatched pairs.
Specifically, it scales logits with a learned temperature parameter as in CoCa, and computes the symmetric cross-entropy loss:

\begin{equation}
    \mathcal{L}_{con} = -\frac{1}{N} \left ( \sum_{i=1}^{N} \log \frac{\exp(\boldsymbol{v}_i^\top \boldsymbol{t}_i/\tau)}{\sum_{j=1}^{N} \exp(\boldsymbol{v}_i^\top \boldsymbol{t}_j/\tau)} + \sum_{i=1}^{N} \log \frac{\exp(\boldsymbol{t}_i^\top \boldsymbol{v}_i/\tau)}{\sum_{j=1}^{N} \exp(\boldsymbol{t}_i^\top \boldsymbol{v}_j/\tau)}\right ),
\end{equation}

where $(\boldsymbol{v}_i, \boldsymbol{t}_i)$ are the vision and language projections respectively, $\tau$ is a temperature parameter learned during training, and $N$ is the batch size.

The report (captioning) objective for report generation minimizes the negative log-likelihood of individual tokens using the factorized joint distribution with teacher-forcing:

\begin{equation}
    \mathcal{L}_{rep} = -\sum_{t=1}^{T} \log p(y_t|y_{<t}, \boldsymbol{X}),
\end{equation} where $y_t | y_{<t}$ is the autoregressive token prediction for token $t$ of $T$ and $\boldsymbol{X}$ is the latent features from the slide encoder.

Besides learning the report generation capability, Yu et al.~\cite{yu2022coca} show that the captioning objective improves instance-level representations compared to using the contrastive objective alone.

The final training objective is a weighted sum of the two the contrastive and generative losses:

\begin{equation}
    \mathcal{L}_{tot} = \lambda_{con}\mathcal{L}_{con} + \lambda_{rep}\mathcal{L}_{rep}.
\end{equation}

\section{Experimental Methods}

\subsection{Training protocols}

We collected a dataset of 587,196 whole slide images corresponding to 195,344 specimens, where each specimen is a collection of one or more \acp{WSI} with a corresponding clinical report.

\begin{figure*}[]
    \centering
    \begin{subfigure}[b]{\textwidth}
         \centering
         \includegraphics{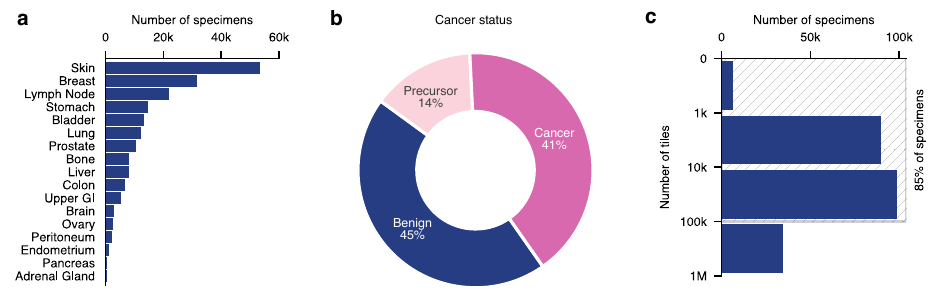}
     \end{subfigure}
    \caption{Statistics on the specimen-level pre-training dataset for PRISM. Note that a specimen may contain one or more \acfp{WSI}. \textbf{a.} Distribution of specimens by the site of tissue origin. \textbf{b.} Proportion of data with the most severe diagnosis being cancer, precursor to caner, or benign. Note for example that a specimen with cancer may also have a precursor to cancer. \textbf{c.} The histogram of tile counts per specimen. 85\% of the specimens (195,344 specimens) have fewer than 100 thousand tiles (plots \textbf{a} and \textbf{b} describe this subset).}
    \label{fig:trainset}
\end{figure*}

Applying our tiling scheme to our dataset results in more than 500,000 tiles for the largest specimen. We constrain our dataset to specimens with up to 100,000 tile embeddings to increase the batch size and speed up training. This constraint does not limit the applicability of the model to clinically-relevant tasks since during application, inference is usually performed on individual slides rather than on the whole specimen and the slides in our dataset have less than 100,000 foreground tiles. See Fig.~\ref{fig:trainset} for the slide and specimen distributions over site of tissue origin and tumor types in the training dataset respectively.

We train the model for 10 epochs on 16 NVIDIA V100 32GB GPU's using fp16 precision with a global batch size of 64 and gradient accumulation over 4 iterations. The model is updated using AdamW with the base learning rate of $2\cdot10^{-4}$ and a cosine decay schedule over 75,000 iterations with 2,000 warm-up iterations. Optimizer weight decay is held fixed at $1\cdot10^{-6}$.

\subsection{Evaluation protocols}

\textbf{Image perception} is performed using either linear probing of fine-tuning.
Linear probing maps the slide embedding from the slide encoder of PRISM (the Perceiver network) to the slide or specimen-level class label.
The Perceiver weights are not updated.
We implement linear probing using logistic regression with 5-fold cross validation and a search over a range of $\ell_2$-regularization coefficients from $1.0$ to $10^6$ in powers of $10$.

To fine-tune the slide encoder we attach a linear classifier as a task head on top of the slide embedding. The classifier's weights are initialized to random values while the slide encoder's weights are pre-trained. We use the same data splits as in linear probing. The Virchow tile-level foundation model is kept frozen.

\textbf{Report generation}. 
PRISM can generate a clinical report for a slide or a specimen using autoregressive decoding.
If tumor is present in the slides, the report describes its type and sub-type, whether it is malignant and/or metastatic, the cancer grade, non-cancerous but relevant features in the surrounding tissue, and the site of tissue origin. In addition, it may predict the results of IHC staining or molecular tests.
Otherwise, the model describes the observed benign tissue.

To generate the reports, the slide encoder (Perceiver) takes a sequence of specimen tile embeddings while the language model is prompted with the special token that indicates the start of a sentence.
The model outputs a probability distribution over its vocabulary and selects the token with the highest output probability (top-1).
The selected token becomes the ground truth for the next token prediction.
The model iteratively decodes the report, halting once the end of sentence token is produced.

\textbf{Zero-shot prediction}.
The contrastive training objective allows for zero-shot evaluation, showcasing how effectively the model can learn clinically relevant tasks directly from diagnostic reports without explicit task-specific training.
It establishes the baseline performance of pre-trained PRISM on novel tasks without further training.

Zero-shot evaluation can be applied to binary classification tasks by creating sets of negative and positive prompts. Specimen slides are processed by the Virchow tile encoder and the slide encoder of PRISM to produce a slide embedding ($\ell_2$-normalized to $\mathbf{v})$, while the language decoder encodes every prompt as a prompt embedding ($\ell_2$-normalized to $\mathbf{t}$), with $\boldsymbol{T}$ being the set of all prompt embeddings. We compute the cosine similarity between each prompt embedding and the slide embedding and apply a softmax activation to the resulting logits with temperature $\tau$ learned during training. Each entry in the vector corresponds to a probability of that prompt matching the slide. In the case of multiple negative and/or positive prompts we perform prompt ensembling by marginalizing probability scores over each set of prompts.
Specifically, a subset of prompt embeddings that belong to class $c$ is defined as $\boldsymbol{T}_c$, and the probability of the class $c$ being the true class given the slide embedding $v$ is

\begin{equation}
    p(C=c \mid \boldsymbol{v}) = \sum_{\boldsymbol{t}_c \in \boldsymbol{T}_c}\frac{\exp(\boldsymbol{v}^\top \boldsymbol{t}_c/\tau)}{\sum_{\boldsymbol{t} \in \boldsymbol{T}} \exp(\boldsymbol{v}^\top \boldsymbol{t}/\tau)}.
\end{equation}

\textbf{Supervised baselines trained from scratch.}
We include a fully-supervised baseline performance for every task to contextualize PRISM's capabilities.
Supervised baselines mirror the fine-tuning protocol except the slide encoder weights are initialized to random values.

\subsection{Evaluation tasks and data} \label{sec:eval_tasks_data}

\textbf{Cancer detection.}
We evaluated cancer detection performance as a binary classification of the presence of cancer in a specimen of multiple \acp{WSI}. We stratified the results in two settings: \textit{All Cancers} and \textit{Rare Cancers} (see~\cite{vorontsov2024virchow}). \textit{All Cancers} contains 22,932 WSIs from 6,142 specimens corresponding to 16 cancer types. The samples are sourced from \ac{MSKCC} (49\%) and other institutions (51\%).  \textit{Rare Cancers} are a subset of \textit{All Cancers} for specimens which meet the National Cancer Institute definition of rare, an incidence of less 15 per 100K per year in the United States. \textit{Rare Cancers} comprises 8753 slides from 2595 specimens corresponding to 7 of the 16 cancer types.

\textbf{Cancer sub-typing.}
We evaluated the cancer sub-typing capabilities of the pre-trained network on three binary tasks: \ac{IDC} versus \ac{ILC} with \ac{TCGA} \ac{BRCA} data, \ac{LUAD} versus \ac{LUSC} with \ac{TCGA} \ac{NSCLC} data, and \ac{DCIS} vs \ac{IDC} on internal \ac{MSKCC} data.

\textbf{Biomarker prediction.}
We evaluated the biomarker prediction performance on 9 different biomarkers originally identified by MSK-IMPACT \cite{cheng2015-msk-impact}, a targeted test for genetic mutations, described in detail in Appendix~\ref{sec:biomarker_appendix} and Table~\ref{tab:supp_biomarker-data-stats}.

Cancer detection and sub-typing with PRISM are evaluated with zero-shot classification, linear probing, and fine-tuning and compared to a baseline slide encoder (Perceiver) trained from scratch. Biomarker prediction is evaluated only with and without slide encoder pre-training (fine-tuning vs supervised).

Most evaluation datasets contain \ac{OOD} data: both Virchow and PRISM were trained only on data internal to \ac{MSKCC}, whereas 49\% of the cancer detection dataset in Table~\ref{tab:all_auc} was composed of cases submitted to \ac{MSKCC} from external sites; data from \ac{TCGA} was not seen during model pre-training. We do not apply stain normalization during evaluation.

\section{Results}

\begin{table*}[t]
  \centering
    \resizebox{\width}{!}{\begin{tabular}{rc|ccc|cc}

    \toprule
            & & \multicolumn{3}{c|}{Cancer sub-typing} & \multicolumn{2}{c}{Cancer detection} \\
            Evaluation & Pre-trained & BRCA$^\dag$ & NSCLC$^\dag$
            & DCIS/IDC$^\ddag$ & Rare Cancers$^\ddag$ & All Cancers$^\ddag$  \\
    \midrule
        Baseline &
            & 0.949 & 0.980 & 0.876 & 0.925 & 0.947
        \\
        Zero-shot & \checkmark
            & 0.952 & 0.975 & 0.908 & 0.868 & 0.906
        \\
        Linear probe & \checkmark
            & \textbf{0.958} & \textbf{0.983} & \textbf{0.939} & 0.931 & 0.949
        \\
        Fine-tuned & \checkmark
            & 0.957 & 0.978 & 0.924 & \textbf{0.938} & \textbf{0.952} 
        \\
    \bottomrule

    \end{tabular}}%
    \caption{
        \Acf{AUROC} performance achieved by PRISM pre-trained slide encoder and compared to a slide encoder baseline trained from scratch on cancer sub-typing and detection tasks (from \acs{TCGA}$^\dag$ or \acs{MSKCC}$^\ddag$). Note that all evaluation methods used pre-trained Virchow embeddings for \ac{WSI} tiles.
    }
  \label{tab:all_auc}%
\end{table*}%

\begin{figure*}[b]
    \centering
    \begin{subfigure}[b]{0.33\textwidth}
         \centering
         \includegraphics[width=\textwidth]{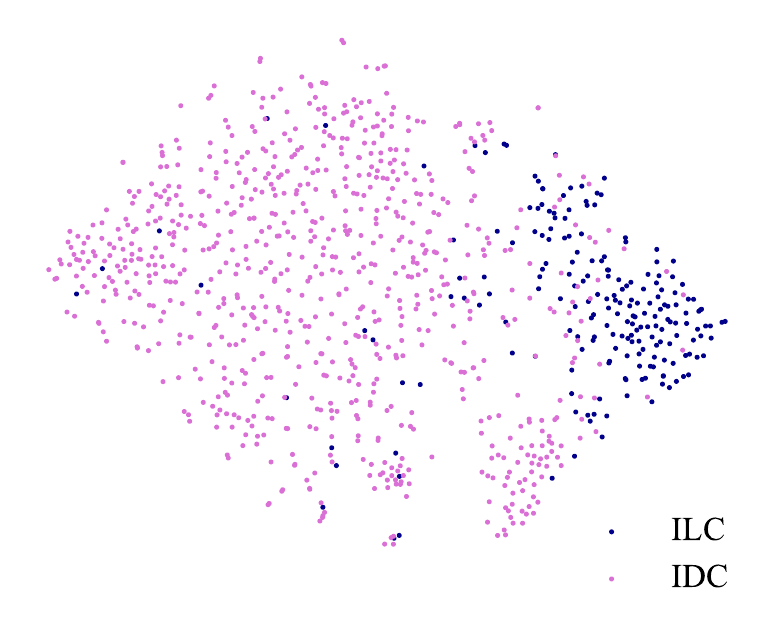}
         \caption{TCGA BRCA}
         \label{fig:brca_tsne}
     \end{subfigure}
     \begin{subfigure}[b]{0.33\textwidth}
         \centering
         \includegraphics[width=\textwidth]{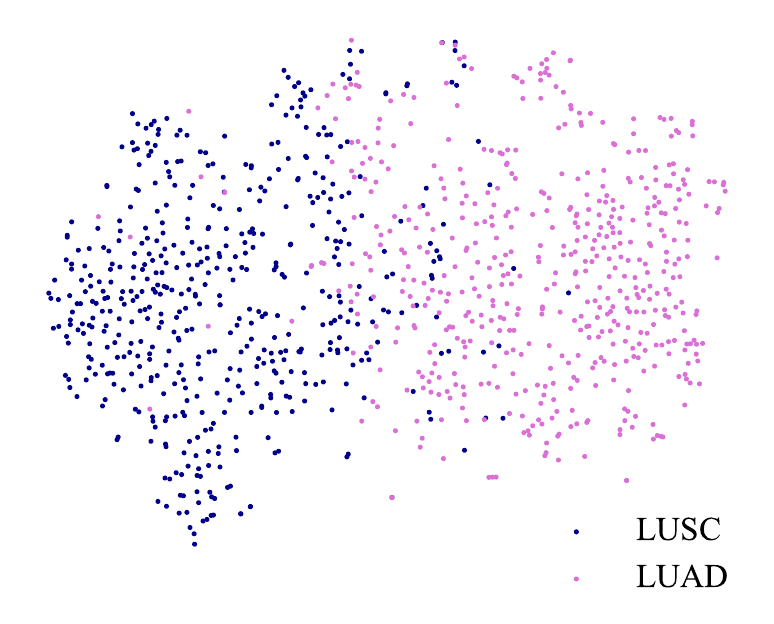}
         \caption{TCGA NSCLC}
         \label{fig:nsclc_tsne}
     \end{subfigure}
     \begin{subfigure}[b]{0.33\textwidth}
         \centering
         \includegraphics[width=\textwidth]{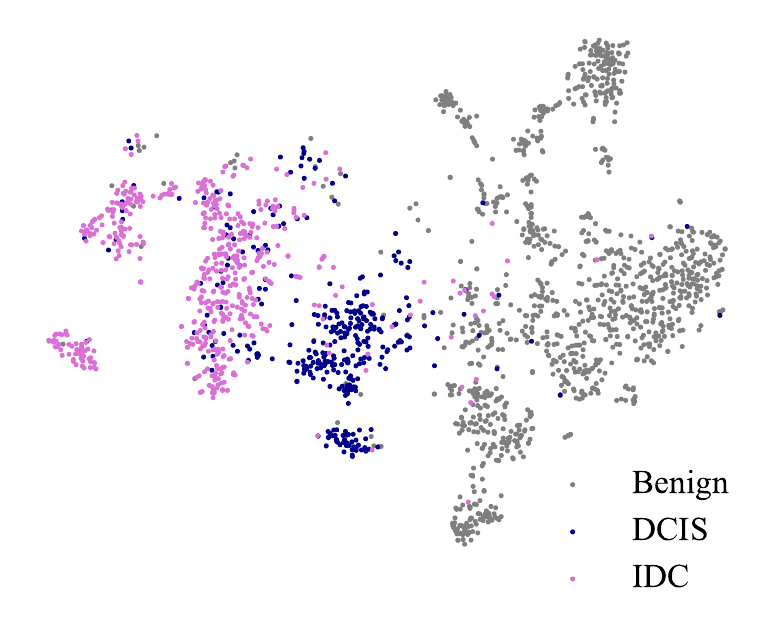}
         \caption{MSK Breast Sub-typing}
         \label{fig:nsclc_tsne}
     \end{subfigure}
    \caption{t-SNE plots of slide embeddings for cancer sub-typing datasets. \Ac{IDC} and \ac{ILC} are types of breast cancer. \Ac{LUSC} and \ac{LUAD} are types of non-small cell lung cancer. \Ac{DCIS} is an early stage non-invasive breast cancer and a precursor to \ac{IDC}; some \ac{IDC} slides can contain \ac{DCIS} regions, however \ac{IDC} takes precedence in the diagnosis as a higher stage cancer. All three plots suggest distinct clusters of slide embeddings in higher dimensions along the cancer sub-types labels.}
    \label{fig:tcga_tsne}
\end{figure*}

\subsection{PRISM demonstrates diagnostic capability}

PRISM pre-trained with pathology reports supervision learns diagnostic capabilities demonstrated on the cancer detection and sub-typing tasks (see Table~\ref{tab:all_auc}). In all cases, performance using the pre-trained model is better than training a model from scratch. Below we describe the results for the zero-shot, linear probing, and fine-tuned settings in detail.

Zero-shot classification performance is comparable to the supervised baseline on the \ac{TCGA} cancer sub-typing tasks: \ac{ILC} vs \ac{IDC} in \ac{BRCA} (0.952 \ac{AUROC} compared to 0.949, Table~\ref{tab:all_auc}) and \ac{LUSC} from \ac{LUAD} in \ac{NSCLC} (0.975 \ac{AUROC} compared to 0.980, Table~\ref{tab:all_auc}). Zero-shot classification outperforms the supervised baseline on \ac{DCIS} vs \ac{IDC} sub-typing in breast on our internal \ac{MSKCC} dataset. These results were achieved without tuning or ensembling text prompts---each class in all sub-typing tasks was encoded using a single prompt selected to resemble the wording in the training data for these sub-types (Table~\ref{tab:prompts}).

On \ac{MSKCC} cancer detection, zero-shot classification performed well but worse than the supervised baseline (0.868 \ac{AUROC} vs 0.925 on rare cancers and 0.906 \ac{AUROC} vs 0.947 on all cancers, Table~\ref{tab:all_auc}). There are a few key differences that likely explain the differences in zero-shot performance across the tasks. First, the cancer detection dataset is much larger than the sub-typing datasets (40,402 specimens in the training set), making it likely that supervised training would perform well. Second, the pan-cancer detection dataset includes 16 cancer types spanning diverse tissue types, 20\% of which were not included in the training set of either Virchow or PRISM. This distribution shift has implications for zero-shot classification as it is possible that some classes have never been observed. Unlike in the cancer sub-typing setting, for cancer detection we ensemble each class over a set of seven prompts (see Tab.~\ref{tab:prompts}).
For the set of positive prompts, we select seven most common cancer types in the pre-training dataset. This is because the reports in the training dataset do not mention ``cancer", instead using more precise terms describing its type, such as lymphoma, sarcoma, adenocarcinoma, and others.
Likewise, the reports don't explicitly state the absence of cancer. Instead, they mention the type of tissue present in the \acp{WSI}, which can include everything from normal tissue described as normal or unremarkable, to benign tumors, inflammation and special cases of inflammation such as Crohn's disease, suspicious or atypical hyperplasia, cancer precursors, and carcinoma in situ. The non-cancer prompts for this task include the seven types of most commonly observed non-cancer histology in the pre-training dataset.
Both sets of prompts do not entirely cover the histology present in the evaluation cancer detection dataset, possibly contributing to reduced performance relative to other evaluation methods. Prompt tuning, using other zero-shot methods like caption scoring, as well as enabling zero-shot cancer detection without the need for ensembling, may improve zero-shot performance and is left as future work.

Linear probing performance surpasses the supervised baseline and zero-shot prediction in all settings. In the cancer sub-typing-tasks, linear probing also outperforms fine-tuning. We use t-SNE to illustrate the separability of the pre-trained slide embeddings of the sub-typing tasks (see Fig.~\ref{fig:tcga_tsne}). Linear probing performance on the \ac{MSKCC} cancer detection is competitive with, but slightly below, the fine-tuned model.

Fine-tuning pre-trained Perceiver leads to better performance than training from scratch and zero-shot and approaches linear probing on the cancer sub-typing tasks. On the \ac{MSKCC} cancer detection task fine-tuning is the best performing method. We suspect the performance gap between linear probe and fine-tuning for sub-typing could be due to overfitting on the relatively small datasets; fine-tuning has superior performance on the much larger \ac{MSKCC} detection dataset. Whole slide aggregation models can be sensitive to the hyperparameters used for training~\cite{bredell2023aggregation} and that hyperparameter tuning could result in better fine-tuning performance.

\begin{figure*}[t]
    \centering
    \includegraphics[width=1.0\textwidth]{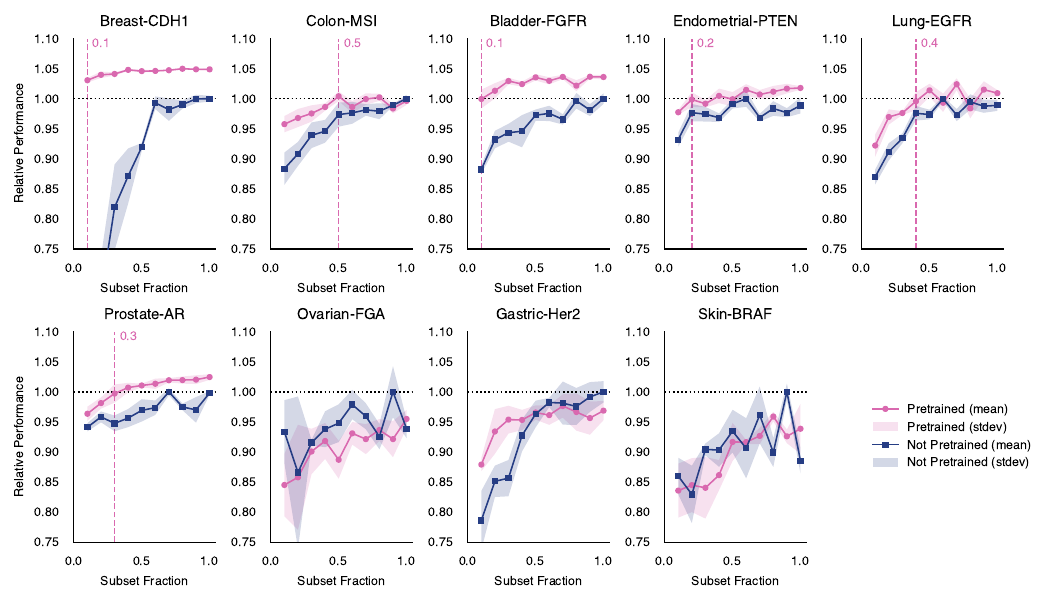}
    \caption{Fine-tuning pre-trained slide encoder (PRISM Perceiver) improves data efficiency compared to training the same model from scratch. The mean (standard deviation) performance across 3 experimental runs is plotted as a solid line (shaded region), relative to the highest \acf{AUROC} achieved without pre-training. The subset fraction denotes the fraction of the available training data used to fine-tune or train the model. A different random subset is selected with each experimental run. The vertical dashed line (magenta) denotes the minimal subset fraction required when fine-tuning PRISM to reach at least 99.5\% of the \ac{AUROC} that can be achieved without pre-training. Note that in many cases, pre-training yields better performance than can be achieved when training from scratch.}
    \label{fig:biomarkers}
\end{figure*}

\subsection{PRISM allows label-efficient training}

Biomarker prediction from routine \ac{HE} stained \ac{WSI} can reduce the delays for patients by removing the need for additional molecular and \ac{IHC} testing. We selected nine biomarkers that play an important role in the diagnosis and treatment of various cancers (see Section ~\ref{sec:biomarker_appendix} for details). The biomarker datasets encode a binary status of genetic alterations (``biomarker'') as measured by DNA extraction and \ac{MSK-IMPACT} sequencing.

Among nine biomarkers, we compare the performance between fine-tuning the pre-trained slide encoder of PRISM and training the same slide encoder from scratch. As shown in Fig.~\ref{fig:biomarkers}, pre-training yielded higher detection performance (\ac{AUROC}) as compared to not pre-training in 6 of the 9 prediction tasks, by as much as 5\% in the case of Breast-CDH1. Furthermore, fine-tuning the pre-trained model was much more label efficient in all 6 tasks. Biomarker prediction was tested with random subsets of each training set in 10\% increments from 10\% to 100\% of the available data. For breast-CDH1 and bladder-FGFR, only 10\% of training data was necessary to meet or exceed the maximum performance attained without pre-training; for endometrial-PTEN, prostate-AR, lung-EGFR, and colon-MSI, 20\%, 30\%, 40\%, and 50\% of the training data was necessary, respectively. Finally, pre-training appears to lower the variance across experiment re-runs for some biomarkers, yielding more consistent results.

\begin{figure*}[p!]
    \centering
    \begin{subfigure}[]{0.8\textwidth}
        \centering
        \includegraphics[width=\textwidth]{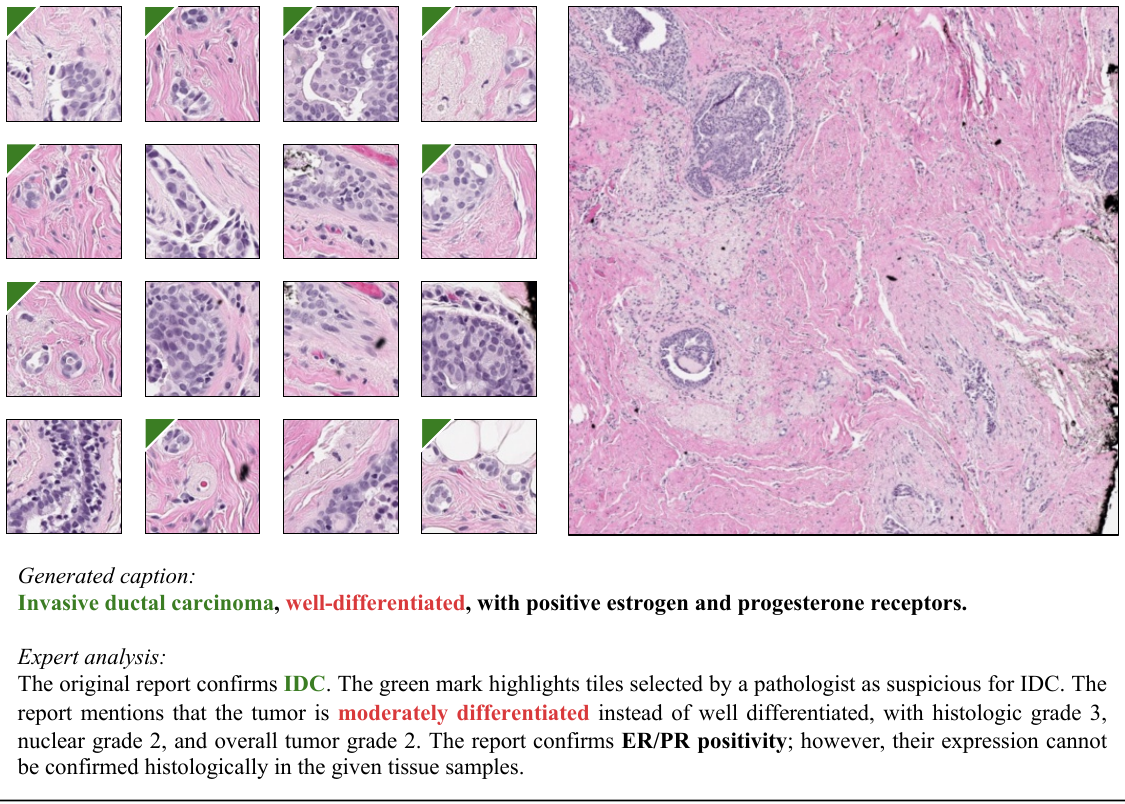}
        \phantomcaption
        \label{fig:attn_idc}
    \end{subfigure}
    \begin{subfigure}[]{0.8\textwidth}
        \centering
        \includegraphics[width=\textwidth]{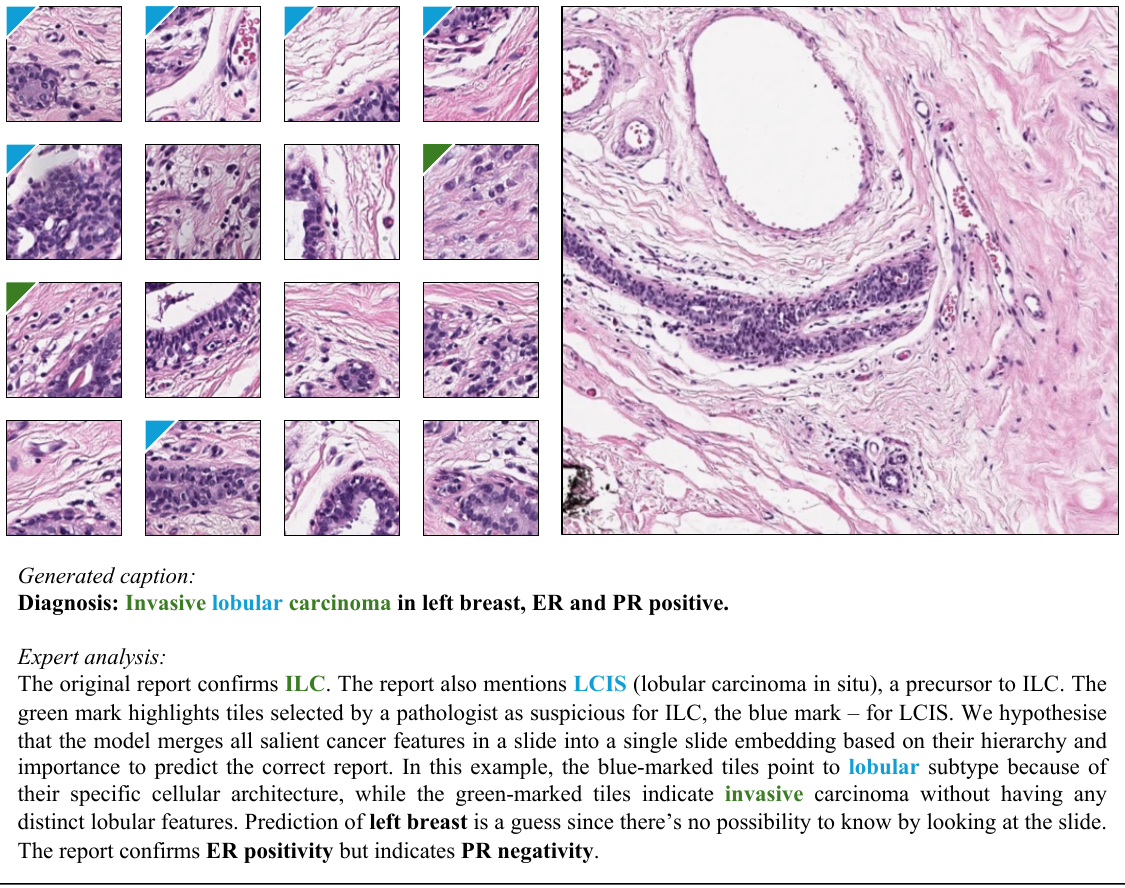}
        \label{fig:attn_ilc}
    \end{subfigure}
\end{figure*}

\begin{figure*}[p!]
    \ContinuedFloat
    \centering
    \begin{subfigure}[]{0.8\textwidth}
        \centering
        \includegraphics[width=\textwidth]{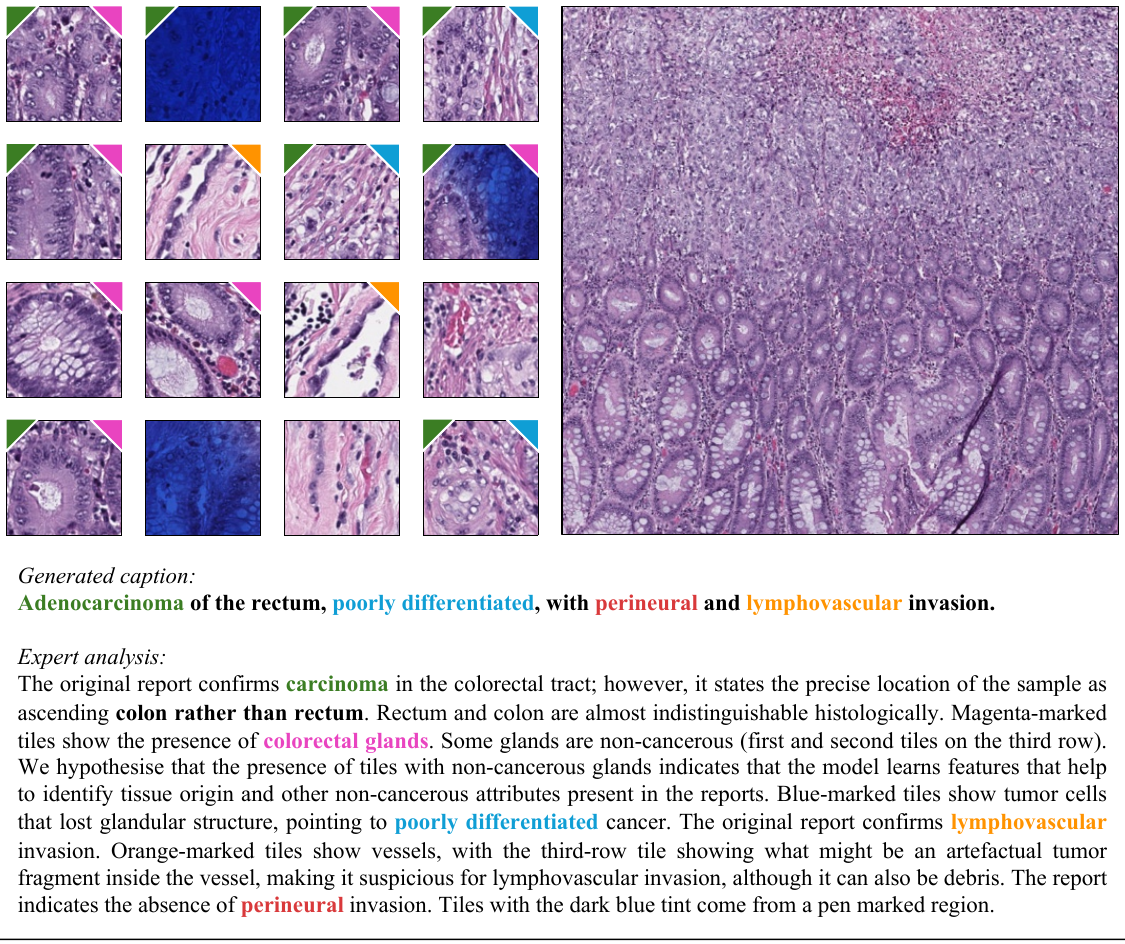}
        \phantomcaption
        \label{fig:attn_colon}
    \end{subfigure}
    \begin{subfigure}[]{0.8\textwidth}
        \centering
        \includegraphics[width=\textwidth]{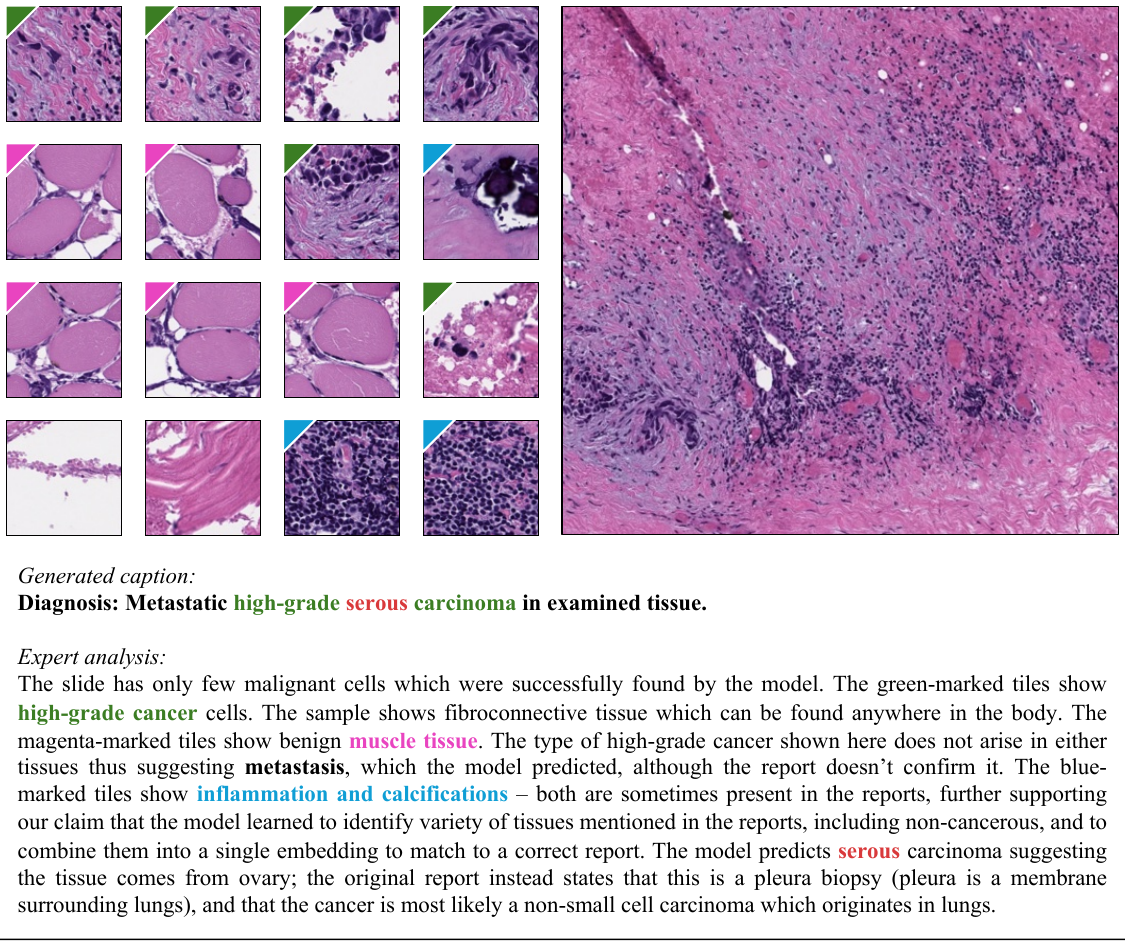}
        \label{fig:attn_pleura}
    \end{subfigure}
\end{figure*}

\begin{figure*}[t!]
    \ContinuedFloat
    \caption{
        Analysis of generated reports and the most attended tiles. The top 16 tiles are shown, weighted by their attention weights in the last cross-attention layer of the Perceiver to the slide embedding latent. The slide embedding is the representation of the \ac{WSI} which learns to summarize all relevant histological features in the slide. Below, a report generated for the corresponding \ac{WSI} and its analysis in relation to the presented tiles. The analysis of the two in conjunction aims to demonstrate that the model learns to select all cancer and non-cancer features in the slide to produce the slide embedding, and at the same time to generated the report that is aligned with the slide embedding. On the right, a region of the \ac{WSI} that contains some of the tiles is shown for context. The image of the region was not used by the model; it is provided only to contextualize model's prediction on the lower resolution level.
    }
    \label{fig:attn}
    \hrule
\end{figure*}

\subsection{PRISM predictions are interpretable}

To contextualize the performance of PRISM and evaluate the explainability of generated reports, we evaluate the 16 tiles most attended by the latent slide embedding. High attention scores (in the last cross-attention layer of the Perceiver slide encoder) indicate that when pooling information across the slide, these tiles have more impact on the final slide embedding than other tiles with lower scores. Four specimens were randomly selected for a thorough review. Fig.~\ref{fig:attn} presents the analysis of the tiles and the corresponding generated reports, performed by a certified pathologist. The pathologist had access to the entire \ac{WSI} at the highest resolution to confirm the diagnosis.

For the given \acp{WSI}, we found a correspondence between the histological features identified by the pathologist in the selected tiles and the text of the generated report. For example, when the generated report mentions \ac{IDC}, many top tiles show invasive cancer cells (see Fig~\ref{fig:attn}, Example 1). Although the pathologist cannot confidently conclude \ac{IDC} in the selected tiles without more context, they can confirm it by looking at the larger region shown on the right. In another example (see Fig~\ref{fig:attn}, Example 3), while the generated report correctly labels the sample as cancerous and many tiles show cancerous tissue, the pathologist identified non-cancerous tissue in some tiles, such as benign colorectal glands.
We hypothesize the presence of these glands among the most attended tiles helps the model to identify the site of tissue origin.
In the generated report, the model predicts rectum as the origin site, which identifies the right organ (the large intestine) but not the correct location (ascending colon), which is histologically indistinguishable from the rectum. 

Overall, the model correctly predicts cancer presence across all four examples and its sub-type in three examples, making a mistake in the example where the sub-type cannot be reliably inferred from the slide alone. It is correct when determining the cancer grade in three examples, making a mistake in one example between a well differentiated and moderately differentiated \ac{IDC}. It sometimes struggles with the site of tissue origin when it cannot be inferred from the slide alone, consequently making a mistake in the sub-type assessment.

\section{Discussion}

Whole slide image analysis is a computationally expensive task that, due to the gigapixel-scale of \acp{WSI}, requires compression of hundreds of thousands of image embeddings containing a vast diversity of features into a single slide embedding. Moreover, typical downstream classification tasks, which include but are not limited to cancer detection, cancer sub-typing, and biomarker prediction, rely on noisy categorical labels which may not provide adequate or robust signals to learn and interpret. Furthermore, tasks such as the prediction of biomarkers or treatment response suffer from an additional challenge of label scarcity. Such factors motivate the design of PRISM, where natural-language clinical reports can be leveraged to learn more nuanced relationships between features and predictions that may also generalize across a wide range of tasks via fine-tuning slide encoder or training a linear classifier on the slide embeddings. Additionally, the ability to generate reports allows for training-free zero-shot classification and enhances the interpretability of the model.

To date, a growing number of works have attempted to learn vision-language pairings for pathology but stayed on the level of tiles. Scaling to the \ac{WSI} level is inhibited by (1) the lack of high-level text descriptions in pre-training datasets and (2) computational bottlenecks of the selected architectures. To address (1), we use \ac{WSI}-level clinical reports summarized by GPT-4~\cite{achiam2023gpt}. To address (2), we use a memory-efficient Perceiver architecture for tile aggregation. PRISM is trained on diverse pathology reports and demonstrates the effectiveness of making predictions on a \ac{WSI} level. Aligning to rich pathology reports boosts in-domain transfer performance as the slide embeddings are encouraged to be adequately diverse, instead of tightly packed about a categorical target, which is illustrated by linear probing and fine-tuning benchmark results. It also benefits transfer to the tasks not covered by the reports (Fig.~\ref{fig:biomarkers}), enabling the types of generalizability that foundation models aim to achieve.

With sufficient data scale, zero-shot and semi-supervised transfer can enable significant improvements in domains where data may be limited, however, it must be noted that a major limitation of this scaling is the availability and quality of clinical reports that cover large variations in the population as well as rare cases not seen during training. Recent research suggests that linear improvements for zero-shot performance requires an exponential increase in multi-modal data for the concepts of interest~\cite{udandarao2024no}. Furthermore, diagnostic reports, while more informative than simple categorical labels, may fail to capture all the necessary details desired in downstream applications and may provide information that is impossible to infer from imaging features derived from \acp{WSI}, e.g. origin site of the tissue sample, introducing noise in the training data. Furthermore, while PRISM appears to perform well on biomarkers that are not represented in the training distribution, evaluation on more tasks is needed to better understand its capabilities. Finally, it has been shown that multi-modal models can over-rely on linguistic priors present in the training data and mostly ignore the information provided by the vision module~\cite{liu2023aligning}. For example, the model may generate a report indicating the presence of a biomarker that is statistically correlated with the input tissue or disease state but not actually present.

There are several areas of this work that warrant future investigation. First, increasing the scale and diversity of the data could result in a better performing and more capable system. Specifically, including richer information about specimens from the diagnostic reports and beyond may allow the model to predict more fine-grained tissue characteristics and understand relationships to molecular and genomic data. This, coupled with further iteration on the GPT-4~\cite{achiam2023gpt} rewriting procedure, could improve downstream capabilities including zero-shot generalization. Another axis of investigation is the scaling of the constituent models. Using a larger and more capable language model should allow for more complex report generation and a better text embedding, while a larger slide embedding model may allow for improved visual features. Moreover, contrastive methods benefit from large and diverse batch sizes ~\cite{radford2021learning} that have not been explored in the current iteration.

Clinical practice has already benefited from computational pathology. Moving towards slide-level models, which more closely match clinical needs, promises to accelerate development even further.

\section{Acknowledgements}
We gratefully thank Philip Rosenfield from Microsoft and Djamilia Dierov from Paige for their contributions in making this collaboration possible, Philippe Mathieu for distributed inference support, Mark Fleishman for data support, and Wayne Hendricks and Alexander van Eck at Paige and Jim Jernigan, Lifeng Li, Ben Huntley, Alex Zhou, Oleg Losinets, and the rest of the team at MSR for infrastructure support.

The results published here are in part based upon data generated by the TCGA Research Network: \url{https://www.cancer.gov/tcga}.

\clearpage
\FloatBarrier
\printbibliography

\onecolumn
\vfill
\pagebreak
\FloatBarrier

\appendix

\renewcommand{\thesection}{A}

\counterwithin{figure}{subsection}
\counterwithin{table}{subsection}

\section{Appendix}

\subsection{Zero-shot prompts}

\begin{table*}[h]
\centering
    \begin{tabular}{p{0.2\textwidth}p{0.14\textwidth}p{0.28\textwidth}p{0.28\textwidth}}
    \toprule
        Task & Classes & Class 1 Prompts & Class 2 Prompts \\
    \midrule
        TCGA BRCA & ILC / IDC & lobular carcinoma, invasive & ductal carcinoma, invasive \\
    \midrule
        TCGA NSCLC & LUSC / LUAD & lung squamous cell carcinoma & lung adenocarcinoma \\
    \midrule
        MSK DCIS/IDC & DCIS / IDC & ductal carcinoma in situ & invasive ductal carcinoma \\
    \midrule
        Pan-cancer detection
        & negative / & benign & cancer \\
        & positive & inflammation & carcinoma \\
        & & normal & adenocarcinoma \\
        & & infectious & malignant \\
        & & cyst & metastatic \\
        & & polyp & invasive \\
        & & unremarkable & sarcoma \\
    \bottomrule
    \end{tabular}%
    \caption{Prompts used in zero-shot evaluations. All tasks are binary classification. The pan-cancer detection dataset uses multiple prompts per class; they are ensembled by taking a sum of probabilities for each prompt in the class to produce the final class score.}
    \label{tab:prompts}%
\end{table*}%

\clearpage
\subsection{Hyperparameters}

\begin{table*}[h]
    \centering
    \begin{tabular}{@{}ll@{}}
    \toprule
    Hyperparameter                          & Value             \\
    \midrule
    Perceiver \\
    Layers                                  & 8                 \\
    Number of latents                       & $512 + 1$         \\
    Latent dimensions                       & $1280$            \\
    Tile embedding dimensions               & $2560$            \\
    Cross-attention \\
    KQV dimensions                          & $1280$            \\
    Heads                                   & $1$               \\
    MLP activation                          & GEGLU             \\
    MLP inner dimensions                    & $1280$            \\
    Layers sharing weights                  & 2-8               \\
    Latent Transformer \\
    Layers                                  & $6$               \\
    KQV dimensions                          & $1280$            \\
    Heads                                   & $8$               \\
    MLP activation                          & GEGLU             \\
    MLP inner dimensions                    & $1280$            \\
    Layers sharing weights                  & 1-8               \\
    \midrule
    BioGPT \\
    Token embedding dimension               & $768$             \\
    Uni-modal layers                        & $1-12$            \\
    Multi-modal layers                      & $13-24$           \\
    \midrule
    Contrastive Projector                    & Linear          \\
    Contrastive embedding dimension          & $5120$          \\
    Contrastive loss weight                  & $1.0$           \\
    Report loss weight                      & $2.0$           \\
    \midrule
    AdamW $\beta$                            & $(0.9, 0.999)$  \\
    Batch size                               & $64$            \\
    Warm up steps                            & $2000$          \\
    Total steps                              & $24000$        \\
    Learning rate schedule                   & Cosine          \\
    Learning rate (start)                    & $2\cdot10^{-4}$          \\
    Weight decay                             & $1\cdot10^{-6}$          \\
    Gradient accumulation iterations         & $4$             \\
    Gradient clipping norm                   & $3.0$           \\
    Automatic mixed precision                & FP16            \\
    \bottomrule
    \end{tabular}%
    \caption{PRISM hyperparameters trained on 16 NVIDIA V100 32GB GPU's.}
    \label{tab:hyperparameters}%
\end{table*}%

\clearpage
\subsection{Pancancer AUCs}
\label{sec:apx_auc}

\begin{table*}[h]
    \centering
    \begin{tabular}{lcccc}
    \toprule
    \multirow{2}{*}{Models} & Perceiver & PRISM & PRISM & PRISM\\
    & (from scratch) & (zero-shot) & (linear probe) & (fine-tuned) \\
    \midrule
        Common origins &                &                &                &                \\
        Breast         & 0.980          & 0.901          & \textbf{0.983} & 0.974          \\
        Prostate       & 0.964          & \textbf{0.975} & 0.929          & 0.944          \\
        Lung           & 0.970          & 0.958          & 0.968          & \textbf{0.975} \\
        Colon          & \textbf{0.998} & 0.991          & 0.996          & 0.996          \\
        Skin           & 0.917          & 0.860          & 0.911          & \textbf{0.939} \\
        Bladder        & 0.933          & 0.943          & \textbf{0.958} & 0.950          \\
        Uterus         & 0.957          & 0.928          & 0.934          & \textbf{0.963} \\
        Pancreas       & \textbf{0.985} & 0.963          & 0.981          & 0.976          \\
        Head and Neck  & 0.987          & 0.966          & 0.985          & \textbf{0.996} \\
    \midrule
        Rare origins   &                &                &                &                \\
        Liver          & \textbf{0.985} & 0.934          & 0.984          & 0.977          \\
        Stomach        & 0.995          & 0.995          & \textbf{0.997} & 0.995          \\
        Brain          & 0.947          & 0.809          & \textbf{0.961} & 0.955          \\
        Ovary          & 0.962          & 0.960          & 0.965          & \textbf{0.969} \\
        Cervix         & 0.853          & 0.870          & 0.852          & \textbf{0.888} \\
        Testis         & 0.952          & 0.904          & 0.901          & \textbf{0.954} \\
        Bone           & 0.800          & 0.683          & 0.818          & \textbf{0.856} \\
    \bottomrule
    \end{tabular}
    \caption{AUC per site of tumor origin in \ac{MSKCC} cancer detection task.}
    \label{tab:apx_pancan_auc}%
\end{table*}%

\clearpage
\subsection{Sub-typing tasks} \label{sec:apx_subtyping_tasks}
\label{sec:sub-typing_appendix}
\textbf{\ac{TCGA} \ac{BRCA}} was curated by selecting only diagnostic slides and then selecting slides whose diagnosis was either \ac{IDC} or \ac{ILC}. Furthermore, we discarded slides that were not readable due to missing metadata or file corruption. This resulted in a dataset with 1002 slides with 798 (80\%) \ac{IDC} and 204 (20\%) \ac{ILC}.

\textbf{\ac{TCGA} \ac{NSCLC}} was curated by selecting diagnostic slides from the \ac{TCGA} \ac{LUAD} and \ac{LUSC} projects. As in \ac{TCGA} \ac{BRCA}, we discarded slides that were not readable due to missing metadata or file corruption. This resulted in a dataset with 1043 slides with 531 (51\%) \ac{LUAD} and 512 (49\%) \ac{LUSC}.

\textbf{\ac{MSKCC} \ac{DCIS}/\ac{IDC}} was curated by selecting from a subset of internal breast specimens that were diagnosed with either \ac{DCIS} or \ac{IDC}. This resulted in a dataset with 896 specimens with 346 (39\%) \ac{DCIS} and 550 (61\%) \ac{IDC}. The dataset includes rare invasive sub-types.

\subsection{Biomarker prediction tasks}
\label{sec:biomarker_appendix}

The training, validation, and testing distribution is shown in Supplementary Tab.~\ref{tab:supp_biomarker-data-stats} for each biomarker dataset. The biomarkers are described below.

\begin{table*}[h]
\centering
\begin{tabular}{@{}rrrrr@{}}
\toprule
Biomarker                         & Subset & Cases & Slides & Positive \\ \midrule
\multirow{3}{*}{Breast-CDH1}      & train  & 648   & 673    & 0.13     \\
                                  & tune   & 215   & 220    & 0.13     \\
                                  & test   & 214   & 228    & 0.13     \\\midrule
\multirow{3}{*}{Colon-MSI}        & train  & 4609  & 11027  & 0.10      \\
                                  & tune   & 481   & 1417   & 0.14     \\
                                  & test   & 482   & 1446   & 0.14     \\\midrule
\multirow{3}{*}{Bladder-FGFR}     & train  & 520   & 542    & 0.24     \\
                                  & tune   & 259   & 275    & 0.29     \\
                                  & test   & 259   & 270    & 0.25     \\\midrule
\multirow{3}{*}{Endometrial-PTEN} & train  & 983   & 1038   & 0.48     \\
                                  & tune   & 164   & 170    & 0.43     \\
                                  & test   & 164   & 178    & 0.41     \\ \midrule
\multirow{3}{*}{Lung-EGFR}        & train  & 2186  & 2858   & 0.28     \\
                                  & tune   & 356   & 457    & 0.29     \\
                                  & test   & 358   & 457    & 0.28     \\\midrule
\multirow{3}{*}{Prostate-AR}      & train  & 1051  & 1461   & 0.18     \\
                                  & tune   & 348   & 480    & 0.20      \\
                                  & test   & 347   & 480    & 0.16     \\ \midrule
\multirow{3}{*}{Ovarian-FGA}      & train  & 679   & 791    & 0.91     \\
                                  & tune   & 115   & 134    & 0.90      \\
                                  & test   & 111   & 126    & 0.88     \\ \midrule
\multirow{3}{*}{Gastric-Her2}     & train  & 968   & 968    & 0.19     \\
                                  & tune   & 170   & 170    & 0.23     \\
                                  & test   & 161   & 161    & 0.17     \\ \midrule
\multirow{3}{*}{Skin-BRAF}        & train  & 782   & 868    & 0.25     \\
                                  & tune   & 131   & 137    & 0.21     \\
                                  & test   & 131   & 138    & 0.13     \\
                                  \bottomrule
\end{tabular}
\caption{Statistics of the case-level biomarker target datasets, including the number of cases (``Cases''), the number of slides (``Slides''), and the proportion of positive labels (``Positive'').}
\label{tab:supp_biomarker-data-stats}%
\end{table*}%

\textbf{Breast-\acs{CDH1}} was assessed via the presence of inactivating mutations associated with \ac{LOH} or a second somatic loss-of-function mutation as determined by \ac{MSK-IMPACT} sequencing test results. Bi-allelic loss of \Acf{CDH1} is strongly correlated with lobular breast cancer, and a distinct histologic phenotype and biologic behavior~\cite{breast_cdh1}. Samples with other types of variants, i.e. mono-allelic mutations, were excluded.

\textbf{Colon-\acs{MSI}} was assessed using both \ac{IHC} and \ac{MSK-IMPACT} sequencing for \ac{dMMR} and \ac{MSI-H} detection, prioritizing \ac{IHC} results when both test outcomes are available. \Ac{MSI-H} is present in approximately 15\% of \acp{CRC}, often linked to germline mutations that elevate hereditary cancer risk. Consequently, routine \ac{MSI} or \ac{IHC}-based \ac{dMMR} screening is recommended for all primary colorectal carcinoma samples. 

\textbf{Bladder-\acs{FGFR}} was assessed via the presence of FGFR3 p.S249C, p.R248C, p.Y373C, p.G370C mutations, FGFR3-TACC3 fusions and FGFR2 p.N549H, pN549K, p.N549S, p.N549T mutations using \ac{MSK-IMPACT} data. The \ac{FGFR} alterations screening in bladder carcinoma allows the identification of patients targetable by \ac{FGFR} inhibitors. 

\textbf{Endometrial-\acs{PTEN}} was assessed via the oncogenic status of \ac{PTEN} mutation using \ac{MSK-IMPACT} data and OncoKB annotation~\cite{chakravarty2017oncokb}. The variants associated with any oncogenic effect (including predicted/likely oncogenic) were defined as positive label for \ac{PTEN} mutations, and variants with unknown oncogenic status were excluded. \ac{PTEN} is the most frequently mutated tumor suppressor gene in endometrial cancer and the presence of \ac{PTEN} mutation has been shown to be significantly associated with poorer prognosis in survival and disease recurrence. 

\textbf{Lung-\acs{EGFR}} was assessed via the oncogenic status of \ac{EGFR} mutation using \ac{MSK-IMPACT} data and OncoKB annotation~\cite{chakravarty2017oncokb}. \Ac{EGFR} mutations with any oncogenic effect (including predicted/likely oncogenic) were defined as positive label, and \ac{EGFR} mutation with unknown oncogenic status were excluded. The \ac{EGFR} oncogenic mutation screening in \ac{NSCLC} is essential to determine eligibility for targeted therapies in late stage \ac{NSCLC}~\cite{egfr_nsclc}. 

\textbf{Prostate-\acs{AR}} was assessed based on whether the fold change of copy number in \ac{AR} was greater than 2 as measured by \ac{MSK-IMPACT} data. \ac{AR} amplification/over-expression is found in 30\%-50\% of \acp{CRPC}, and is associated with the resistance to \ac{ADT}.

\textbf{Ovarian-\acs{FGA}} was assessed based on whether the fraction of genome altered was $\geq$ 30\% as measured by \ac{MSK-IMPACT} data. This cutoff has been established to enrich for TP53 mutations, a factor in the characterization of \Ac{HGSOC} and previously reported to correlate to increased FGA.

\textbf{Gastric-\acs{HER2}} amplification was assessed based on \ac{HER2} \ac{IHC} 2+/FISH+ or \ac{IHC} 3+. Approximate 20\% of gastric cancer patients are found to correlate with \ac{HER2} overexpression / high-level amplification and would likely benefit from treatment with an anti-\ac{HER2} antibody therapy.

\textbf{Skin-\acs{BRAF}} was assessed based on  oncogenic mutation status and the presence of V600E variant using \ac{MSK-IMPACT} data and OncoKB annotation~\cite{chakravarty2017oncokb}. \ac{BRAF} is one of the most frequently mutated genes in melanoma, and V600E mutation is the most common variant, which leads to constitutive activation of the BRAF/MEK/ERK (MAPK) signalling pathway. Targeted therapy with \ac{BRAF} inhibitors showed better survival outcome in patients with \ac{BRAF} V600-mutated melanoma.  

\subsection{Biomarker results}
\begin{table*}[h]
    \centering
    \begin{tabular}{lcccc}
    \toprule
    \multirow{2}{*}{Biomarker} & \multicolumn{2}{c}{Not pre-trained} & \multicolumn{2}{c}{Pre-trained} \\
    & AUC & f & AUC & f \\
    \midrule
    Breast-CDH1 & 0.938 (0.009) & 1.0 & \textbf{0.984} (0.004) & 0.8 \\
    Colon-MSI & 0.966 (0.012) & 1.0 & \textbf{0.969} (0.015) & 0.5 \\
    Bladder-FGFR & 0.847 (0.015) & 1.0 & \textbf{0.878} (0.005) & 0.9 \\
    Endometrial-PTEN & 0.833 (0.023) & 0.6 & \textbf{0.847} (0.005) & 1.0 \\
    Lung-EGFR & 0.770 (0.006) & 0.6 & \textbf{0.789} (0.015) & 0.7 \\
    Prostate-AR & 0.800 (0.010) & 0.7 & \textbf{0.820} (0.007) & 1.0 \\
    Ovarian-FGA & \textbf{0.825} (0.070) & 0.9 & 0.788 (0.025) & 1.0 \\
    Gastric-Her2 & \textbf{0.812} (0.029) & 1.0 & 0.793 (0.032) & 0.7 \\
    Skin-BRAF & \textbf{0.736} (0.020) & 0.9 & 0.706 (0.012) & 0.8 \\
    \bottomrule
    \end{tabular}
    \caption{The top \acf{AUROC} achieved for various biomarker tasks when training the tile aggregator with and without pre-training with PRISM, shown as mean (standard deviation) across 3 experimental runs. The fraction f of the training set used to produce this result is also shown.}
    \label{tab:biomarker_results}%
\end{table*}%

\clearpage
\subsection{GPT-4 prompt for clinical report rewriting} \label{sec:apx_gpt4_rewrites}

\begin{table*}[h]
    \centering
    \begin{tabular}{@{}ll@{}}
    \toprule
    Parameter                               & Value              \\
    \midrule
    Model                                   & gpt-4-1106-preview \\
    Temperature                             & 0.7                \\
    Top P                                   & 0.95               \\
    \bottomrule
    \end{tabular}
    \caption{API parameters for GPT-4 report rewriting}
    \label{tab:gpt4-api-parameters}%
\end{table*}%

System prompt:

\begin{tcolorbox}[breakable,boxrule=0pt]
  You are a pathology lab assistant. You are given an unstructured pathology report describing a tissue sample. Follow these instructions carefully and to the letter:\\
- Extract a detailed summary of the diagnosis and the examined tissue from the report in a sentence under 20 words.\\
- Mention tissue type if it is mentioned in the report.\\
- If the report includes results of immunohistochemical studies or molecular tests, include a short summary of the most important positive results. Keep the overall sentence length below 20 words.\\
- Do not mention any negative test results or absence of something.\\
- Do not mention any numbers.\\
- Do not mention cm or mm measurements.\\
- Minimize prose, be succinct, use as few words as possible.\\
- Following the instructions above, rewrite the report 5 times, for a total of 5 sentences, each with a summary written in a slightly different way. Make sure all summaries are consistent with each other.\\
- Output the resulting 5 sentences as a list in JSON format. Don't output anything else.
\end{tcolorbox}

\hfill

Examples of rewritten reports:

\begin{tcolorbox}[breakable,boxrule=0pt]
Metastatic urothelial carcinoma in one lymph node with extranodal extension. \\
One lymph node shows metastatic urothelial carcinoma with extranodal spread. \\
Diagnosis: Urothelial carcinoma metastasis in lymph node, extranodal extension present. \\
Lymph node involved by metastatic urothelial carcinoma; extranodal extension observed. \\
Urothelial carcinoma with metastasis to a lymph node and extranodal extension identified.
\end{tcolorbox}

\begin{tcolorbox}[breakable,boxrule=0pt]
Diagnosed cutaneous neuroendocrine Merkel cell carcinoma with lymphovascular invasion. \\
Merkel cell carcinoma of the skin with infiltrative growth and lymphovascular invasion found. \\
Skin Merkel cell carcinoma confirmed, showing infiltrative pattern and lymphovascular invasion. \\
Cutaneous Merkel cell carcinoma with lymphovascular invasion, infiltrative pattern diagnosed. \\
Pathology reveals skin Merkel cell carcinoma, lymphovascular invasion present.
\end{tcolorbox}

\begin{tcolorbox}[breakable,boxrule=0pt]
Acral skin biopsy; reactive changes, no melanoma, PRAME positive, SOX10 positive. \\
Examined acral skin; found reactive changes, PRAME and SOX10 immunopositivity, no melanoma. \\
Acral skin showing reactive changes; melanoma absent; positive for PRAME and SOX10. \\
Reactive changes in acral skin tissue; PRAME and SOX10 positive; melanoma not detected. \\
Acral skin with reactive changes; tests positive for PRAME and SOX10; no evidence of melanoma.
\end{tcolorbox}

\begin{tcolorbox}[breakable,boxrule=0pt]
Benign breast parenchyma diagnosed. \\
Diagnosis: Benign condition of breast tissue. \\
Breast tissue exam reveals benign parenchyma. \\
Benign parenchymal findings in breast tissue. \\
Examined breast tissue is benign.
\end{tcolorbox}

\clearpage
\subsection{Examples of generated reports}

\begin{table*}[ht!]
\centering
\begin{tabular}{p{0.22\textwidth}p{0.72\textwidth}}
\toprule
Relevant report sections & Left ovary and fallopian tube:
- Left ovary with follicular cysts and focal dystrophic calcification.
- Left fallopian tube with tubal cysts.
- No tumor seen.
\\
Generated report & Benign ovarian and fallopian tube tissue with calcifications.
\\
Expert comment & Correct diagnosis and site
\\

\midrule
Relevant report sections & Core biopsy right distal femur:  
- Enchondroma
\\
Generated report & Diagnosis: Enchondroma in bone tissue.
\\
Expert comment & Correct diagnosis and site
\\

\midrule
Relevant report sections & Liver, core biopsy:  
- Hepatocellular carcinoma, moderately differentiated, see note
Note:  Submitted reticulin stain supports the diagnosis.  Submitted immunohistochemical stain for glypican 3 is focally positive in tumor cells.
\\
Generated report & Liver biopsy: Well-differentiated hepatocellular carcinoma identified.
\\
Expert comment & Correct diagnosis and site
\\

\midrule
Relevant report sections & Prostate, right medial base, mid, apex:   
a.
Prostate biopsy location: Right Base medial
Benign Processes: Benign prostatic tissue, Acute inflammation, Chronic inflammation
b.
Prostate biopsy location: Right Mid medial
Benign Processes: Benign prostatic tissue, Acute inflammation, Chronic inflammation
c.
Prostate biopsy location: Right Apex medial
Benign Processes: Benign fibromuscular stroma; no prostatic glands are identified
\\
Generated report & Diagnosis: Benign prostatic tissue with acute and chronic inflammation.
\\
Expert comment & Correct diagnosis and site
\\

\midrule
Relevant report sections & Esophagus and proximal stomach (fs):
Tumor Type: Invasive adenocarcinoma
Treatment effect (Status Post neoadjuvant therapy): Treatment response involves 80\% of the tumor
Histologic Grade: Moderately differentiated
Tumor Location: Lower thoracic portion esophagus ($<$ 5cm above EGJ) extends 1\% into the stomach
Siewert Type: Type I ($>$ 75\% above GEJ)
Tumor Size: Length is 1.5 cm, Width is 1.4 cm, Maximal thickness is 0.5 cm
Depth of Tumor Invasion: Tumor invades the muscularis propria, Foci of acellular mucin likely representing regressed tumor are present in the subserosal tissue.
Vascular Invasion: Not identified
Perineural Invasion: Not identified
Tumor Multicentricity: Not identified
Adjacent Mucosa: Unremarkable squamous mucosa, Gastric mucosa shows focal intestinal metaplasia, negative for dysplasia
Proximal Margin: Free of carcinoma or dysplasia, with 4.5 cm clearance
Distal Margin: Free of carcinoma or dysplasia
Deep Margin: Free of invasive carcinoma with a clearence of 0.5 cm
Lymph Nodes: Number of nodes examined: 17, Number of metastatic nodes: 0
Tumor Stage (AJCC 7th Edition): pT2 (Invades the muscularis propria)
Lymph Node Stage (AJCC 7th Edition): pN0 (No regional node metastasis)
\\
Generated report & Diagnosis: Metastatic adenocarcinoma in a lymph node.
\\
Expert comment & Diagnosis is correct. Discordance: no lymph nodes are present in the specimen.
\\

\midrule
\end{tabular}%
\phantomcaption
\end{table*}%
\begin{table*}[ht!]
\centering
\begin{tabular}{p{0.22\textwidth}p{0.72\textwidth}}
\midrule
Relevant report sections & Left partial glossectomy, floor of mouth resection:
Procedure: Resection, Other: Partial glossectomy
Histologic Type: Squamous cell carcinoma, keratinizing
Histologic Grade: G1: Well differentiated
Tumor Location: Floor of mouth
Tumor Laterality: Left
Tumor Focality: Single focus
Tumor Size: Greatest dimension is 2.9 cm (per report)
Tumor Thickness: Tumor thickness: 8 mm, Depth of invasion: 4.8 mm
Microscopic Tumor Extension: Invades skeletal muscle
In situ carcinoma: Not Identified
Specimen Margins:  Margins involved by invasive tumor (specify margin): Deep margin
Tumor Bed Margins: Not applicable
Lymph-Vascular Invasion: Not identified
Perineural Invasion: Not identified
Lymph Nodes: No lymph nodes submitted or found
Extranodal Extension (ENE): Not applicable
Additional Pathologic Findings: None identified
\\
Generated report & Diagnosis: Well-differentiated squamous cell carcinoma with perineural invasion in the dermis.
\\
Expert comment & Diagnosis and site are correct. Perineural invasion is in caption but not identified in the ground truth.
\\

\midrule
Relevant report sections & Rectum, 10 cm, biopsy:  
- Invasive adenocarcinoma, moderately differentiated, see note
Note: Immunohistochemical stains for MMR proteins will be reported in an
addendum.
\\
Generated report & Moderately differentiated invasive adenocarcinoma in rectum; MLH1, MSH2, MSH6, PMS2 proteins retained.
\\
Expert comment & Diagnosis and site are correct. MMR biomarkers are not reported in original report provided but it would be of interest to know if all were retained (common).
\\

\midrule
Relevant report sections & Lung, superior segment LLL; segmentectomy:
Tumor Location: Lung, left lower lobe, Superior segment
Procedure: Segmentectomy
Tumor Type: Adenocarcinoma
Predominant pattern: Micropapillary 60\%
Other patterns: Acinar  30\%, Solid 10\%, The acinar component includes 10\% cribriform pattern
Tumor Focality: Unifocal
Tumor Size: Total tumor size 1.8 cm
Histologic Grade: Poorly differentiated (G3)
Lymphovascular Invasion: Present (extensive)
Perineural Invasion: Not identified
Spread Through Air Spaces: Present
Pleural Invasion: The tumor invades into the visceral pleura beyond the elastic
layer (PL1)
Bronchial Margin: For final margin status, please see part 3
Vascular Margin: For final margin status, please see part 3
Staple Line: For final margin status, please see part 3
Distance from margin: Cannot be assessed, See part 3
Peribronchial or intrapulmonary Lymph Nodes: Number with metastasis: 1
\\
Generated report & Adenocarcinoma with mucinous features in lung; CK7, CK20, CDX2 positive.
\\
Expert comment & Partially correct: site and tumor are correct, however no mention of any mucinous component.
\\

\midrule
Relevant report sections & Skin and soft tissue, right leg; wide excision:  
- Metastatic melanoma.  
\\
Generated report & Diagnosis: Metastatic melanoma in lymph node with extranodal extension. 
\\
Expert comment & Morphological diagnosis is correct however the report does not mention any lymph nodes.
\\

\midrule
Relevant report sections & Left hemiglossectomy:   
Procedure: Glossectomy (hemiglossectomy)
Histologic Type: Squamous cell carcinoma, keratinizing
Histologic Grade: G2: Moderately differentiated
Tumor Location: Tongue
Tumor Laterality: Left
Tumor Focality: Single focus
Tumor Size: Greatest dimension is 5.0 cm per outside report
Tumor Thickness: Depth of invasion 18 mm, per outside report
Microscopic Tumor Extension: Invades skeletal muscle
In situ carcinoma: Not Identified in the submitted material
Specimen Margins: Cannot be assessed, Location of closest margin, per orientation: cannot be assessed in the submitted material
Tumor Bed Margins: Not applicable
Lymph-Vascular Invasion: Not identified in the submitted material
Perineural Invasion: Not identified in the submitted material
Lymph Nodes: No lymph nodes submitted or found in the submitted material
\\
Generated report & Lymph node with metastatic keratinizing squamous carcinoma; extranodal extension present.
\\
Expert comment & Morphological diagnosis is correct however the report mentions no lymph nodes submitted.
\\

\midrule
\end{tabular}%
\phantomcaption
\end{table*}%
\begin{table*}[ht!]
\centering
\begin{tabular}{p{0.22\textwidth}p{0.72\textwidth}}
\midrule
Relevant report sections & Right mandible, biopsy:  
Small fragment of benign skeletal muscle, fibrous tissue, bone, and dentin. No epithelium-lined cyst seen
\\
Generated report & Bone and muscle tissue show benign pathology.
\\
Expert comment & Correct, all benign findings.
\\

\midrule
Relevant report sections & Breast, left lateral calcifications; core biopsy:
In Situ Carcinoma: Markedly atypical ductal hyperplasia focally reaching the level of
ductal carcinoma in situ (DCIS)
DCIS, Architectural Pattern: Cribriform
Nuclear Grade: Low
Non-neoplastic Breast Tissue: A few foci suspicious for atypical lobular hyperplasia (ALH)
Calcifications: Rare, in DCIS and in stroma
\\
Generated report & Diagnosis: Ductal carcinoma in situ (DCIS) in breast tissue with cribriform pattern.
\\
Expert comment & Correct diagnosis of DCIS.
\\

\midrule
Relevant report sections & Skin, left chest: 
Procedure: Excisional biopsy
Tumor Type: Melanoma
Histologic Type: Superficial spreading type
Breslow thickness: 0.95 mm
Ulceration: Not identified
Mitotic index: 1 /mm2
Clark level: IV
Growth phase: Vertical
Infiltrating lymphocytes: Non-brisk
Regression: identified, involving less than 75\% of lesion
Lymphovascular Invasion: Not identified
Perineural Invasion: Not identified
Microscopic satellite: Not identified
Associated nevus: Not identified
Surgical Margins: Not involved, The distance of melanoma in situ from the nearest side margin is:1mm.
Tumor Staging: pT1b (The melanoma is less than or equal to 1 mm thick. It is
ulcerated and/or the mitotic rate is equal to or greater than 1/mm2)
\\
Generated report & Diagnosis: Superficial spreading melanoma, Clark level IV, with ulceration.
\\
Expert comment & Correct diagnosis, minor discordance in one part where the GT says no ulceration but it describes ulceration. Otherwise diagnosis and grade are correct.
\\

\midrule
Relevant report sections & Right Fallopian Tube and Ovary, Excision/Resection:  
Serous cystadenoma (4.5 cm)
Fallopian tube within normal limits
\\
Generated report & Diagnosis: Serous cystadenoma of the ovary, fallopian tube unremarkable.
\\
Expert comment & Correct diagnosis and site.
\\

\midrule
Relevant report sections & Uterus, cervix, bilateral tubes and ovaries; total hysterectomy and
bilateral salpingo-oophorectomy:
- Atrophic endometrium with focal glandular crowding
- Benign endometrial polyp
- Adenomyosis
- Benign ovaries with simple cysts
- Benign fallopian tubes with paratubal cysts
\\
Generated report & Diagnosis: Endometrioid adenocarcinoma, FIGO grade 1, with myometrial invasion.
\\
Expert comment & Incorrect diagnosis, major discordance.
\\

\midrule
Relevant report sections & Ileal mucosa with no significant pathologic abnormalities.
\\
Generated report & Small lymphocytic lymphoma / chronic lymphocytic leukemia identified in lymphoid tissue.
\\
Expert comment & Incorrect diagnosis, major discordance.
\\

\midrule
Relevant report sections & Distal pancreas for tumor bank
Specimen(s) Received: Pancreas, tail
Procedure: Distal pancreatectomy
Histologic Type: Residual adenocarcinoma exhibiting changes consistent with
treatment effect (status post neoadjuvant therapy)
Histologic Grade: Poorly differentiated
Tumor Location: Tail of the pancreas
Overall Tumor Size: Greatest dimension: 5.0 cm, Per outside report
Microscopic Tumor Extension: Tumor invades peripancreatic soft tissues
Lymph-Vascular invasion: Present
Perineural Invasion: Present
Margins: Uninvolved by invasive carcinoma/pre-invasive neoplasm per report
Anterior Peritoneal Surface: Uninvolved by invasive carcinoma
Posterior Peritoneal Surface: Free of invasive carcinoma/pre-invasive neoplasm
Histologic Evidence of Treatment Effect: Present, Involves approximately 60
Adjacent Pancreas: Exhibits atrophy, Exhibits fibrosis
Lymph Nodes: Number of lymph nodes examined in this specimen: 16, Number of lymph nodes with metastasis in this specimen: 2
TNM descriptors: y (posttreatment)
Primary Tumor (pT) (AJCC 8th Edition): pT3: Tumor $>$4 cm in greatest dimension
Regional Lymph Nodes (pN) (AJCC 8th Edition): pN1: Metastasis in one to three regional lymph nodes
\\
Generated report & Metastatic high-grade serous carcinoma found in lymph node tissue.
\\
Expert comment & Metastatic carcinoma is correct however saying “serous carcinoma” is specific to mullerian origin (e.g., ovary) and thus incorrect.
\\

\midrule
\end{tabular}%
\phantomcaption
\end{table*}%
\begin{table*}[ht!]
\centering
\begin{tabular}{p{0.22\textwidth}p{0.72\textwidth}}
\midrule
Relevant report sections & Gastric cancer. History of met gastric signet ring cell cancer.
Pylorus, biopsies:
Invasive adenocarcinoma, poorly differentiated.
\\
Generated report & Diagnosis: Poorly differentiated adenocarcinoma with signet ring features in stomach.
\\
Expert comment & Correct diagnosis and site.
\\

\midrule
Relevant report sections & Benign cystic tissue consistent with clinical impression of hydrocele.
\\
Generated report & Examined tissue: Benign ovarian cyst.
\\
Expert comment & Benign diagnosis is correct, incorrect site of origin.
\\

\midrule
Relevant report sections & Breast, left; lumpectomy:   
Invasive Carcinoma: Mucinous
Size:  Microscopic measurement 1.5 cm
Histologic Grade: III/III: minimal or no tubule formation ($<$ 10\% of tumor) [score 3]
Nuclear Grade: II/III (moderate variation in size and shape) [score 2]
Mitotic Count: $<$ 8 mitoses per 10 high power fields [score 1]
Overall Tumor Grade: Combined score 6: grade II/III (moderately differentiated)
Ductal Carcinoma in Situ (DCIS): Identified, Extends into lobules
DCIS Architectural Pattern: Solid, Cribriform
DCIS Nuclear Grade: Intermediate
Calcification: In DCIS, In benign epithelium
Lymphovascular Invasion: Not identified
Surgical Margins (Invasive Carcinoma): For final margin status see separately submitted margins
Surgical Margins (DCIS): For final margin status see separately submitted margins
Non-neoplastic Breast Tissue: Biopsy site changes
\\
Generated report & Breast tissue shows mucinous carcinoma, Grade II-III, with mucin production.
\\
Expert comment & Correct diagnosis and site.
\\

\bottomrule
\end{tabular}%
\caption{We compare relevant report sections of the original clinical reports with the reports generated by PRISM. The reports are on specimen level, many spanning multiple \acp{WSI}. The specimens come from the evaluation cancer detection dataset (Section~\ref{sec:eval_tasks_data}.}
\label{tab:generated_reports}%
\end{table*}%

\clearpage
\subsection{Perceiver implementation details}
\label{sec:apx_perceiver}

The cross-attention module takes tile embeddings as key-value tokens and resamples them with latent tokens as queries using single-head attention~\cite{vaswani2017attention}.
Layer normalization~\cite{ba2016layer} is applied to latents but not to tile embeddings in order to save memory.
The latent self-attention transformer is a 6-layer transformer encoder that closely follows Vaswani et al.~\cite{vaswani2017attention}, with 8 heads per attention module.
The MLP networks in the cross-attention modules and latent transformers consist of a GEGLU layer~\cite{shazeer2020glu} followed by a linear layer, with 1280 inner dimensions.
We don't apply position encoding to tile embeddings.

\begin{figure}[hp!]
\begin{lstlisting}[language=Python, caption={Perceiver slide encoder pseudocode}, label={lst:code}]

def perceiver(
    latents: Tensor,  # learned latent embeddings (513, 1280)
    embeddings: Tensor,  # Virchow tile embeddings (N, 2560)
    xattn_layer_0: nn.Module,  # cross-attention module with MLP
    xattn_layer_1: nn.Module,  # cross-attention module with MLP
    latent_transformer: nn.Module,  # 6-layer transformer
    perceiver_depth: int = 8,
) -> tuple[Tensor, Tensor]:  # return slide embedding and latent features
    for i in range(perceiver_depth):
        # Cross-attention
        if i == 0:
            latents, kv_cache = xattn_layer_0(
                latents, context=embeddings, kv_cache=None
            )
        elif i == 1:
            latents, kv_cache = xattn_layer_1(
                latents, context=embeddings, kv_cache=None
            )
        else:
            # Layers 2 through 8 share the weights with layer 1
            # and take KV cache instead of tile embeddings
            latents, kv_cache = xattn_layer_1(
                latents, context=None, kv_cache=kv_cache
            )

        # Latent transformer; layers 1 through 8 share the weights
        latents = latent_transformer(latents)

    return (
        latents[0, :],  # slide embedding (1280,)
        latents[1:, :],  # latent features to language decoder (512, 1280)
    )

\end{lstlisting}
\end{figure}

\begin{figure}[hp!]
\begin{lstlisting}[language=Python, caption={Cross-attention module pseudocode}, label={lst:code}]

def xattn_layer_i(
    latents: Tensor,  # learned latent embeddings (513, 1280)
    context: Tensor | None,  # Tile embeddings (N, 2560), or None if kv_cache
    kv_cache: Tensor | None,  # Keys and values (2, N, 1280), or None if context
    attention: nn.Module,  # QKV-attention with W_q, W_k, W_v, W_o linear layers
    layer_norm_1: nn.Module,  # Layer normalization layer before attention
    mlp: nn.Module,  # 2-layer MLP with GEGLU non-linearity and inner dim 1280
    layer_norm_2: nn.Module,  # Layer normalization layer before MLP
) -> tuple[Tensor, Tensor]:  # return new latents and new kv_cache
    output, kv_cache = attention(q=layer_norm_1(latents), kv=context or kv_cache)
    latents = latents + output
    latents = latents + mlp(layer_norm_2(latents))
    return latents, kv_cache

\end{lstlisting}
\end{figure}

\clearpage
\subsection{Acronyms}
\begin{acronym}[MSK-IMPACT]
    \acro{ADT}{androgen deprivation therapy}
    \acro{AI}{artificial intelligence}
    \acro{AR}{androgen receptor}
    \acro{AUROC}[AUC]{area under (the receiver operating characteristic) curve}
    \acro{BRAF}{B-Raf Proto-Oncogene}
    \acro{BRCA}{breast cancer}
    \acro{CDH1}{cadherin 1}
    \acro{CNN}{convolutional neural network}
    \acro{CRC}{colorectal cancer}
    \acro{CRPC}{castration resistant prostate cancer}
    \acro{dMMR}{deficient mismatch repair}
    \acro{EGFR}{epidermal growth factor receptor}
    \acro{FGA}{fraction of genome altered}
    \acrodefplural{FGA}{fractions of genome altered}
    \acro{FGFR}{fibroblast growth factor receptor}
    \acro{HER2}{human epidermal growth factor receptor 2}
    \acro{HE}[H\&E]{hematoxylin and eosin}
    \acro{HGSOC}{high-grade serous ovarian cancer}
    \acro{HIPT}{the hierarchical image pyramid transformer}
    \acro{IHC}{immunohistochemistry}
    \acro{LOH}{loss-of-heterozygosity}
    \acro{MIL}{multiple instance learning}
    \acro{MMR}{mismatch repair}
    \acro{mpp}{microns-per-pixel}
    \acro{MSI-H}{high-frequency MSI}
    \acro{MSI}{microsatellite instability}
    \acro{MSK-IMPACT}{MSK-Integrated Mutation Profiling of Actionable Targets}
    \acro{MSKCC}{Memorial Sloan Kettering Cancer Center}
    \acro{MTC}{medullary thyroid cancer}
    \acro{NSCLC}{non-small cell lung cancer}
    \acro{OOD}{out-of-distribution}
    \acro{PTEN}{phosphatase and tensin homolog}
    \acro{RET}{Ret Proto-Oncogene}
    \acro{RNN}{recurrant neural network}
    \acro{TCGA}{The Cancer Genome Atlas}
    \acro{ViT}{vision transformer}
    \acro{WSI}{whole slide image}
    \acro{LUSC}{lung squamous cell carcinoma}
    \acro{LUAD}{lung adenocarcinoma}
    \acro{IDC}{invasive ductal carcinoma}
    \acro{ILC}{invasive lobular carcinoma}
    \acro{DCIS}{ductal carcinoma in situ}
    \acro{LCIS}{lobular carcinoma in situ}
\end{acronym}

\end{document}